%% file: equivalence.tex
\documentclass[9pt, twocolumn]{IEEEtran}
\IEEEoverridecommandlockouts

\usepackage{amsthm}
\usepackage{enumerate}
\usepackage{ifthen}
\usepackage{verbatim}
\usepackage{multirow}
\usepackage{multicol}

\usepackage{cleveref}

\usepackage{cite}
\usepackage[usenames,dvipsnames]{color}

\usepackage{graphicx,amssymb,amstext,amsmath,amsfonts}

\usepackage{balance}
\usepackage{xcolor}
\usepackage{multirow}
\usepackage{lscape}
\usepackage{mdwlist}
\usepackage{url}
\usepackage{subfigure}

\usepackage[font=normal, center]{caption}

\usepackage{dcolumn}
\usepackage{mathrsfs}
\usepackage{latexsym}

\usepackage[latin1]{inputenc}
\usepackage{tikz}
\usetikzlibrary{shapes,arrows}

\DeclareMathOperator{\rank}{\text{rank}}

\newboolean{showcomments}
\setboolean{showcomments}{true}
\newcommand{\bose}[1]{\ifthenelse{\boolean{showcomments}}
{ \textcolor{blue}{(Bose says: #1)} } {} }
\newcommand{\slow}[1]{\ifthenelse{\boolean{showcomments}}
{ \textcolor{red}{(Steven says:  #1)}}{}}
\newcommand{\mani}[1]{\ifthenelse{\boolean{showcomments}}
{ \textcolor{green}{(Mani says:  #1)}}{}}

\graphicspath{{figures/}}

\tikzstyle{decision} = [diamond, draw,
    text width=4.5em, text badly centered, node distance=3cm, inner sep=0pt]
\tikzstyle{block} = [rectangle, draw,
    text width=5em, text centered, rounded corners, minimum height=4em]
\tikzstyle{line} = [draw, -latex']
\tikzstyle{cloud} = [draw, ellipse, node distance=3cm,
    minimum height=2em]

\makeatletter
\newtheorem*{rep@theorem}{\rep@title}
\newcommand{\newreptheorem}[2]{%
\newenvironment{rep#1}[1]{%
 \def\rep@title{#2 \ref{##1}}%
 \begin{rep@theorem}}%
 {\end{rep@theorem}}}
\makeatother

\newtheorem{theorem}{Theorem}
\newtheorem{lemma}[theorem]{Lemma}
\newtheorem{corollary}[theorem]{Corollary}

\newtheorem{remark}{Remark}

\def\ba{\begin{array}}
\def\ea{\end{array}}
\newcommand{\beq}{\begin{equation}}
\newcommand{\eeq}{\end{equation}}
\newcommand{\bq}{\begin{eqnarray}}
\newcommand{\eq}{\end{eqnarray}}
\newcommand{\bqn}{\begin{eqnarray*}}
\newcommand{\eqn}{\end{eqnarray*}}
\newcommand{\bee}{\begin{enumerate}}
\newcommand{\eee}{\end{enumerate}}
\newcommand{\bi}{\begin{itemize}}
\newcommand{\ei}{\end{itemize}}
\newcommand{\btab}{\begin{tabular}}
\newcommand{\etab}{\end{tabular}}
\newcommand{\ii}{\textbf{i}}

\newcommand{\PP}{\mathcal{P}}
\newcommand{\RR}{\mathcal{R}}
\newcommand{\FF}{\mathbb{W}}
\newcommand{\XX}{\mathbb{X}}

\newcommand{\herm}{{H}}
\renewcommand\Re{\text{Re }}
\renewcommand\Im{\text{Im }}
\renewcommand{\Cref}{\eqref}
\renewcommand{\cref}{\eqref}

\allowdisplaybreaks[1]

\begin{document}

\title{Equivalent relaxations of optimal power flow
\thanks{A preliminary and abridged version has appeared in \cite{Bose-2012-BFMe-Allerton}.}
}
\author{
Subhonmesh Bose$^{1}$, Steven H. Low$^{2, 1}$, Thanchanok Teeraratkul$^{1}$, Babak Hassibi$^{1}$ \\
$^{1}$Electrical Engineering, $^{2}$Computing and Mathematical Sciences\\
California Institute of Technology	\\
}
\maketitle

\begin{abstract}
Several convex relaxations of the optimal power flow (OPF) problem have recently been developed using both bus injection models and branch flow models. In this paper, we prove relations among three convex relaxations: a semidefinite relaxation that computes a full matrix, a chordal relaxation based on a chordal extension of the network graph, and a second-order cone relaxation that computes the smallest partial matrix. We prove a bijection between the feasible sets of the OPF in the bus injection model and the branch flow model, establishing the equivalence of these two models and their second-order cone relaxations. Our results imply that, for radial networks, all these relaxations are equivalent and one should always solve the second-order cone relaxation. For mesh networks, the semidefinite relaxation is tighter than the second-order cone relaxation but requires a heavier computational effort, and the chordal relaxation strikes a good balance. Simulations are used to illustrate these results.
\end{abstract}


\input{introBose.tex}
\input{BusInjection2.tex}

\input{BranchFlow2.tex}

\input{modelEquiv2.tex}

\input{numerics2.tex}

\section{Conclusion}
\label{sec:conc}
In this paper, we have presented various conic relaxations of the OPF problem and their relations in both
the bus injection and the branch flow models. In the bus injection model 
the SDP relaxations $\RR_1$ and $\RR_{ch}$ are equivalent and are generally 
tighter than the SOCP relaxation $\RR_2$. For acyclic  networks however 
these relaxations are equivalent. The branch flow model leads to an SOCP  
relaxation $\tilde \RR_2$. We have shown that $\RR_2$ and $\tilde \RR_2$ are equivalent. 
In general $\RR_{ch}$ is faster to compute than $\RR_1$. 
$\RR_2$ and $\tilde \RR_2$ are even faster, though their feasible sets are generally
larger than that of $\RR_1$ or $\RR_{ch}$.

\section*{Acknowledgment}
We are thankful to Prof. K. Mani Chandy and Lingwen Gan
  at Caltech for helpful discussions.  We also acknowledge the support of NSF through NetSE grant CNS 0911041, DoE's
ARPA-E through grant DE-AR0000226,  the National Science Council
of Taiwan (R. O. C.) through grant NSC 103-3113-P-008-001,
Southern California Edison, and the Resnick Institute at Caltech.

\bibliographystyle{unsrt}
\balance
\bibliography{../../../PowerRef-201202}

\end{document}

%% file: introBose.tex
\section{Introduction}
\label{sec:intro}

\subsection{Background}

The optimal power flow (OPF) problem seeks an operating point of a power network
that minimizes a certain cost,
 e.g., generation cost, transmission losses, etc.   It is a fundamental problem as it underlies many
applications such as unit commitment, economic dispatch, state estimation, volt/var
control, and demand response.   There has been a great deal of research since 
Carpentier's first formulation in 1962 \cite{Carpentier62} and an early solution
by Dommel and Tinney \cite{Dommel1968}; recent surveys can be found in, e.g.,
\cite{Powerbk, Huneault91,Momoh99a,Momoh99b,Pandya08, Frank2012a, Frank2012b,
OPF-FERC-1, OPF-FERC-2, OPF-FERC-3, OPF-FERC-4, OPF-FERC-5, Low2013}.
OPF is generally nonconvex and NP-hard.   A large
number of optimization algorithms and relaxations have been proposed,
the most popular of which is linearization (called DC OPF)
\cite{Stott1974,Alsac1990,Purchala2005,Stott2009};
See also \cite{Coffrin2012} for a more accurate linear approximation.
An important observation was made in \cite{bai2008} that  OPF can be
formulated as a quadratically constrained quadratic program and therefore can be
approximated by a semidefinite program (SDP).  
Instead of solving  OPF  directly, the authors in \cite{Lavaei2012} propose to 
solve its convex Lagrangian dual problem.
Sufficient conditions have been studied by many authors under which an optimal solution for the non-convex problem can be derived from an optimal solution of its SDP relaxation; e.g., \cite{Bose2011, Zhang2011geometry, Sojoudi2012PES} for radial networks
and in \cite{Lavaei2012, Rantzer2011, Gan-2013-OPFDC} for resistive networks.
These papers all use the standard bus injection model where the Kirchhoff's
laws are expressed in terms of the complex nodal voltages in rectangular 
coordinates.

Branch flow models on the other hand formulate OPF in terms of branch 
power  and current flows in addition to nodal voltages, e.g., 
\cite{Baran1989a, Baran1989b, Cespedes1990, Exposito1999, Jabr2006, 
Farivar2011-VAR-SGC, Taylor2012-TPS, Jabr2012}.
They have been mainly used for modeling radial distribution networks.
A branch flow model has been proposed in \cite{Farivar-2013-BFM-TPS} to study
OPF for both radial and mesh networks and a relaxation based on second-order cone program (SOCP) is developed.
Sufficient conditions are obtained in \cite{Farivar2011-VAR-SGC, Gan-2012-BFMt, Li-2012-BFMt}
under which the SOCP relaxation is exact for radial networks.

\subsection{Summary}

Since the OPF problem in the bus injection model 
is a quadratically constrained quadratic program it is equivalent to 
a rank-constrained SDP \cite{bai2008, Lavaei2012}.   This formulation naturally leads to
an SDP relaxation that removes the rank constraint and solves for a full
positive semidefinite matrix.
If the rank condition is satisfied at an optimal point, the relaxation is said to be \emph{exact} and an optimal solution of OPF can be recovered through the spectral decomposition of the positive semidefinite matrix.
Even though SDP is polynomial time solvable it is nonetheless impractical 
to compute for large power networks.   Practical networks, however, are sparse.
In this paper we develop two equivalent formulations of OPF
using {\em partial matrices} that involve much fewer variables than the
full SDP.

The key idea is to characterize classes of partial matrices that are easy
to compute and, when the relaxations are exact, are completable to full positive 
semidefinite matrices of rank 1 from which a
solution of OPF can be recovered through spectral decomposition.
One of these equivalent problems leads to an SDP relaxation based on 
chordal extension of the network graph \cite{Fukuda99exploitingsparsity, klerk2010}
and the other leads to an SOCP relaxation \cite{Lobo1998,Boyd2004}.
In this work, we prove equivalence relations among these problems and their relaxations.  
Our results imply that, for radial networks, all three relaxations are equivalent and 
we should always solve the SOCP relaxation.  
For mesh networks there is a tradeoff between computational effort and
accuracy (in terms of exactness of relaxation) in deciding between 
solving SOCP relaxation or the other two 
relaxations.   Between the chordal relaxation and the full SDP, 
 if all the maximal cliques of a chordal extension of the network graph
have been pre-computed offline then solving the chordal relaxation
is always better because it has the same accuracy as the full 
SDP  but typically involves far fewer
variables and is faster to compute.
This is explained in Section \ref{sec:bim}.
Chordal relaxation has been suggested in \cite{Bai2011, Jabr2012} for
solving OPF, and SOCP relaxation in the bus injection model has also been studied in 
\cite{Sojoudi2012PES, Bose-2012-QCQPt, Bose-2012-BFMe-Allerton, Gan-2013-OPFDC}.
Here we provide a framework that unifies and contrasts these approaches.

In Section \ref{sec:bfm} we present the branch flow model of 
\cite{Farivar-2013-BFM-TPS} for OPF and the corresponding SOCP relaxation developed in
\cite{Farivar2011-VAR-SGC, Farivar-2013-BFM-TPS}.  
In Section \ref{sec:eq} we prove the equivalence of the branch flow model and
the bus injection model by exhibiting a bijection between these two models
and their relaxations.
Indeed the relations among the various problems in this paper, both in the bus
injection model and the branch flow model, are established through relations among
their feasible sets.

It is important that we utilize both the bus injection and the branch flow models.   
Even though they are equivalent, some relaxations are
much easier to formulate and some sufficient conditions for exact relaxation are 
much easier to prove in one model than the other.  
For instance the semidefinite relaxation of power flows has a much cleaner 
formulation in the bus injection model.  
The branch flow model especially for radial networks has a convenient 
recursive structure that not only allows a more efficient computation of power flows
e.g. \cite{Kersting2002, Shirmohammadi1988, ChiangBaran1990}, but
also plays a crucial role in proving the sufficient conditions for exact relaxation 
in \cite{Gan-2013-BFMt-CDC, Gan-2013-BFMt-TAC}.
Since the variables in
the branch flow model correspond directly to physical quantities such as
branch power flows and injections it is sometimes more convenient in
applications.

In Section \ref{sec:numerics}, we illustrate the relations among the various relaxations and OPF through simulations. First, we visualize the feasible sets of a 3-bus example
in \cite{Lesieutre-2011-OPFSDP-Allerton}. Then we compare the running times and accuracies of these relaxations on IEEE benchmark systems \cite{UW_data, Zimmerman09}.
We conclude the paper in Section \ref{sec:conc}.

\subsection{Notations}

 Let $\mathbb{R}$ and $\mathbb{C}$ denote the sets of real and complex numbers 
 respectively.  
For vectors $x, y \in \mathbb R^n$, $x\leq y$ denotes inequality componentwise;
if $x, y\in \mathbb C^n$, $x\leq y$ means $\Re x \leq \Re y$ and $\Im x \leq \Im y$.
For a matrix $A$, let $A^H$ be its hermitian transpose.
$A$ is called \emph{positive semidefinite (psd)}, denoted  $A \succeq 0$, if it is
hermitian and $x^H A x \geq 0$ for all $x\in \mathbb C^n$.  
Let $\ii := \sqrt{-1}$ and for any set $B$, let $| B |$ denote its cardinality.


%% file: BusInjection2.tex
\section{Bus injection model and conic relaxations} 
\label{sec:bim}

In this section we formulate OPF in the bus injection model and describe three
equivalent problems.   These problems lead naturally to semidefinite relaxation,
chordal relaxation, and second-order cone relaxation of OPF.   We prove
equivalence relations among these problems and their exact relaxations.

\subsection{OPF formulation}

Consider a power network  modeled by a connected undirected graph 
$G(N, E)$ where each node in $N := \{1, 2, \ldots, n \}$ represents a bus 
and each edge in $E$ represents a line.
For each edge $(i,j)\in E$
let $y_{ij}$ be its admittance \cite{Bergen2000}.
A bus $j \in N$ can have a generator, a load, both or neither.  Typically the
loads are specified and the generations 
are variables to be determined.   Let $s_j$ be the net complex power injection
(generation minus load) at bus $j\in N$. Also, let $V_j$ be the complex voltage at bus $j \in N$ and $|V_j|$ denote its magnitude. Bus 1 is the slack bus with a fixed magnitude $|V_1|$ (normalized to 1).  
The \emph{bus injection model} is defined by the following power flow equations
that describe the Kirchhoff's law\footnote{The current flowing from bus $j$ to bus $k$ is $(V_j - V_k) y_{jk}$.}:
\bq
s_j  & = & \sum_{k: (j, k) \in E} V_j (V_j^H - V_k^H) y_{jk}^H
\ \ \ \text{ for } j \in N.
\label{eq:opf.1a}
\eq
The power injections at all buses satisfy
\bq
 \underline{s}_j  \leq s_j \leq \overline{s}_j \ \ \ \ \text{ for } j \in N,
 \label{eq:opf.2a}
\eq
where $\underline{s}_j$ and $\overline{s}_j$ are known limits on the net injection
at bus $k$. 
It is often assumed that the slack bus (node 1) has a generator and there is no
limit of $s_1$; in this case $-\underline{s}_j = \overline{s}_j = \infty$.
We can eliminate the variables $s_k$ from the OPF formulation by combining
\eqref{eq:opf.1a}--\eqref{eq:opf.2a} into
\bq
\underline{s}_j  \ \ \leq  \sum_{k: (j, k) \in E} V_j (V_j^H - V_k^H) y_{jk}^H \ \ 
\leq \ \ \overline{s}_j
\ \ \ \text{ for } j \in N.
\label{eq:opf.1}
\eq
Then OPF in the bus injection model can be formulated in terms of just the $n \times 1$ voltage
vector $V$.    All voltage magnitudes are  constrained:
 \bq
\underline{V}_j \leq |V_j| \leq \overline{V}_j  \quad \text{ for } j \in N,
  \label{eq:opf.2}
\eq
where $\underline{V}_j$ and $\overline{V}_j$ are known lower and upper voltage limits. 
Typically $|V_1| = 1 = \underline{V}_1 = \overline{V}_1$.
These constraints define  the feasible set of the optimal power flow problem in the bus
injection model:
\begin{align}
\mathbb{V} := \{ V \in \mathbb{C}^n \ \vert \ V \text{ satisfies } \eqref{eq:opf.1}-\eqref{eq:opf.2} \}.
\label{eq:opf.4}
\end{align}

Let the cost function be $c(V)$.  Typical costs include the total cost of
generating real power at all buses or line loss over the network.
All these costs can be expressed as  functions of $V$.  
Thus, we obtain the following optimization problem.
\\
\noindent
\textbf{Optimal power flow problem $OPF$:}
\begin{subequations}
\label{OPF}
\begin{align*}
& \underset{V}{\text{minimize}}   \quad\  c(V) \\
& \text{subject to} \quad\   V \in \mathbb V.
\end{align*}
\end{subequations}
Since \Cref{eq:opf.1} is quadratic, $\mathbb{V}$ is generally a nonconvex set. Thus
OPF is nonconvex and NP-hard to solve.

\begin{remark}
The OPF formulation usually includes additional constraints such as thermal or stability limits
on power or current flows on the lines, or security constraints; see surveys in
\cite{Powerbk, Huneault91,Momoh99a,Momoh99b,Pandya08, OPF-FERC-1, 
OPF-FERC-2, OPF-FERC-3, OPF-FERC-4, OPF-FERC-5}.
Our results generalize to OPF with some of these  constraints, e.g., line limits
\cite{Bose-2012-QCQPt, Farivar-2013-BFM-TPS}.
Our model can also include a shunt element at each bus.  We omit these refinements
 for  ease of presentation.
\end{remark}

\subsection{SDP relaxation: $\PP_1$ and $\RR_1$}

\begin{figure*}[t]
\centering
	\subfigure[Graph $G_1$]{ { \scalebox{0.5}{\includegraphics*{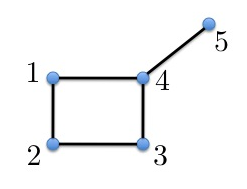}}} \label{fig:graph1} }
	\subfigure[Graph $G_2$]{ {\scalebox{0.5}{\includegraphics*{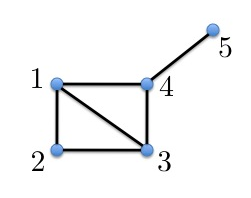}}} \label{fig:graph2}}
	\caption{Simple graphs to illustrate $G$-partial matrices.} \label{fig:graphs}
\centering
	\subfigure[$G_1$-partial matrix]{ { \scalebox{0.5}{\includegraphics*{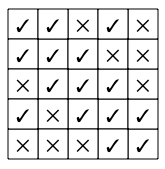}}} \label{fig:graph1.mat} } \hspace{2.5cm}
	\subfigure[$G_2$-partial matrix]{ {\scalebox{0.5}{\includegraphics*{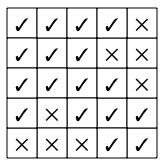}}} \label{fig:graph2.mat}}
	\caption{Index sets $I_{G_1}$ and $I_{G_2}$ illustrated as entries in a matrix. Entry $(j, k)$ is marked with a tick if $(j, k)$ is in the corresponding index set; otherwise it is marked with a cross.} \label{fig:graphs.mat}
\end{figure*}

Note that \cref{eq:opf.1} is linear in the variables $W_{jj} := |V_j|^2$ for 
$j \in N$ and $W_{jk} := V_j V_k^H$ for $(j, k) \in E$.
This motivates the definition of a $G$-partial matrix. 
Define the index set $I_G$:
\begin{align*}
I_G := \Bigg\{(j,j) \ \vert \ j \in N \Bigg\} \ \bigcup \ \Bigg\{ (j,k) \ \vert \  (j, k) \in  {E} \Bigg\}.
\end{align*}
A \emph{${G}$-partial matrix} $W_G$ is a collection of complex numbers 
indexed by the set $I_G$, i.e., $[W_G]_{jk}$ is defined iff $j=k \in N$ or $(j, k) \in E$. 
This is illustrated in Figure \ref{fig:graphs}. For graph $G_1$, we have $n=5$ nodes 
and $I_{G_1} = \{ (1,1), (2,2), (3,3), (4,4), (5,5), (1,2), (2,1), (2,3),  \linebreak[0] (3,2), (3,4), (4,3), (1,4), (4,1), (4,5), (5,4)  \}$
 as shown in Figure \ref{fig:graph1.mat} as a partially filled matrix. 
 For graph $G_2$ in Figure \ref{fig:graph2}, $I_{G_2}$ is represented in Figure 
 \ref{fig:graph2.mat}.  If $G$ is a {\em complete graph}, i.e., every pair of nodes share an edge, then $W_G$
 is an $n \times n$ matrix. 

The relations in \eqref{eq:opf.1}--\eqref{eq:opf.2} can be rewritten in terms of $W_G$ as:
\begin{subequations}
\bq
\underline{s}_j  \leq  \sum_{k: (j,k)\in E} \left( [W_G]_{jj} - [W_G]_{jk} \right) y_{jk}^H
 \leq  \overline{s}_j    \; \text{ for } j \in N,
\label{eq:opfW.1}
\\
\underline{V}_j^2 \leq [W_G]_{jj} \leq \overline{V}_j^2 \; \text{ for } j \in N.
\label{eq:opfW.2}
\eq
\end{subequations}

We assume the cost function $c(V)$ in OPF depends on $V$ only through the
$G$-partial matrix $W_G$.   For instance, 
if the objective is to minimize the total real power loss in the network then 
$$c(V) = \sum_{j \in N} \Re s_j = \sum_{j \in N} \sum_{k:(j,k)\in E} 
\Re \left([W_G]_{jj} - [W_G]_{jk} \right) y_{jk}^H.$$
If the objective is to minimize a weighted sum of real power generation at various nodes then
\begin{align*}
c(V) &= \sum_{j \in N} c_j \left(\Re s_j - p_j^{d} \right)\\
&= \sum_{j \in N} c_j  \left( \sum_{k:(j,k)\in E} 
\Re \left([W_G]_{jj} - [W_G]_{jk} \right) y_{jk}^H - p_j^D \right),
\end{align*}
where $p_j^{d}$ is the given real power demand at bus $j \in N$. 
Henceforth we refer to the cost function as $c(W_G)$.

Consider an $n \times 1$ voltage vector $V$. Then $W = V V^H$ is an $n \times n$ psd matrix of rank 1. Define the $G$-partial matrix $W(G)$ as the collection of $I_G$ entries of $W$. To describe the constraints $V \in \mathbb V$, we use the equivalent constraints in terms of $W(G)$ in \eqref{eq:opfW.1}-\eqref{eq:opfW.2}. Formally, OPF is equivalent to the following problem with $n \times n$ Hermitian matrix $W$:
\\
\textbf{Problem $\PP_1$:}
\bqn
\underset{{W}}{\text{minimize}}  & & c(W(G)) \\
\text{subject to} & & W(G) \text{ satisfies } \eqref{eq:opfW.1}-\eqref{eq:opfW.2},\\ 
& & {W} \succeq 0, \  \rank W = 1.
\eqn
Given an $V\in \mathbb V$, $W = VV^H$ is feasible for $\PP_1$; conversely given a feasible
$W$ it has a unique spectral decomposition \cite{horn05} $W = VV^H$ such that $V\in \mathbb V$.  Hence
there is a one-one correspondence between the feasible sets of OPF and $\PP_1$, i.e., 
OPF is equivalent to $\PP_1$. Problem $\PP_1$ is a rank-constrained SDP and NP-hard to solve.
The nonconvex rank constraint is relaxed to obtain the following SDP.\\
\textbf{Problem $\RR_1$:}
\bqn
\underset{{W}}{\text{minimize}}  & & c(W(G)) \\
\text{subject to} & & W(G) \text{ satisfies } \eqref{eq:opfW.1}-\eqref{eq:opfW.2}, 
\quad {W} \succeq 0.
\eqn
$\RR_1$ is an SDP \cite{wolkowicz00, Boyd2004} and can be solved in polynomial time using interior-point algorithms \cite{nesterov1987interior, alizadeh1995interior}. Let $W^*$ be an optimal solution of $\RR_1$. If $W^*$ is rank-1 then $W^*$ also solves $\PP_1$ optimally. We say the relaxation \emph{$\RR_1$ is exact with respect to $\PP_1$} if there exists an optimal solution of $\RR_1$ that satisfies the rank constraint in $\PP_1$ and hence optimal for $\PP_1$. 
\begin{remark}
\label{rem:exactRel}
In this paper we define a relaxation to be exact as long as one of its optimal
solutions satisfies the constraints of the original problem, even though a relaxation
may have multiple optimal solutions with possibly different ranks.
The exactness of $\RR_1$ in general does not guarantee that we can compute efficiently a rank-1 
optimal $W_*$ if non-rank-1 optimal solutions also exist.
Many sufficient conditions for exact relaxation in the recent literature, however, do
guarantee that \emph{every} optimal solution of the relaxation is optimal for the 
original problem, e.g., 
\cite{Zhang2013, LavaeiTseZhang2012, LamZhang2012, Gan-2013-OPFDC} 
or they lead to a polynomial time algorithm to construct an optimal solution of $\PP_1$ from 
\emph{any} optimal solution of the relaxation, e.g., \cite{Sojoudi2013, Bose-2012-QCQPt}.
\end{remark}


\subsection{Chordal relaxation: $\PP_{ch}$ and $\RR_{ch}$}

To define the next relaxation we need to extend  the definitions  of
 Hermitian, psd, and rank-1 for matrices to partial matrices:
 \bee
\item 
The complex conjugate transpose
of a $G$-partial matrix $W_{G}$ is the $G$-partial matrix $(W_G)^H$ that satisfies
\begin{align*}
[(W_G)^H]_{jk} = [W_G]_{kj}^H  \text{ for all } (j, k) \in I_G.
\end{align*}
We say $W_G$ is  \emph{Hermitian} if $W_G = (W_G)^H$.

\item 
 A matrix $M$ is psd if and only if all its principal submatrices (including
 $M$ itself) are psd.   We extend the definition of psd to $G$-partial matrices using this 
 property.  Informally a $G$-partial matrix is said to be psd if, when viewed as a partially
 filled $n\times n$ matrix, all its \emph{fully-specified} principal submatrices are psd.
 This notion can be formalized as follows.
  A clique is a complete subgraph of a given graph. A clique on $k$ nodes is 
   referred to as a $k$-clique. For the graph $G_1$ in Figure \ref{fig:graph1}, the cliques are the edges. 
   For the graph $G_2$ in Figure \ref{fig:graph2}, the cliques consist of the edges and the triangles 
   $\{ 1, 2, 3 \}$ and $\{1, 3, 4 \}$. 
A $k$-clique $C$ in graph $G$ on nodes $\{ n_1, n_2, \ldots, n_k \}$ fully specifies the 
$k\times k$ submatrix $W_G (C)$\footnote{For any graph $F$, a partial matrix $W_F$, and a subgraph $H$ of $F$, the partial matrix $W_F(H)$ is a submatrix
of $W_F$ corresponding to the $I_H$ entries of $W_F$. If subgraph $H$ is a $k$ clique, then $W_F(H)$ is a $k\times k$ matrix.}:
\begin{align*}
W_G(C) = \begin{pmatrix}
  [W_G]_{n_1 n_1} & [W_G]_{n_1 n_2} & \cdots & [W_G]_{n_1 n_k} \\
  [W_G]_{n_2 n_1} & [W_G]_{n_2 n_2} & \cdots & [W_G]_{n_2 n_k} \\
  \vdots  & \vdots  & \ddots & \vdots  \\
  [W_G]_{n_k n_1} & [W_G]_{n_k n_2} & \cdots & [W_G]_{n_k n_k}
 \end{pmatrix}.
\end{align*}
We say a $G$-partial matrix $W_G$ is {\em positive semidefinite (psd)}, 
written as $W_G \succeq 0$,
if and only if $W_G(C) \succeq 0$ for all cliques $C$ in graph $G$.

\item
A matrix $M$ has rank one  if $M$ has exactly one linearly independent row (or column). 
We say a $G$-partial matrix $W_G$ has rank one, written as $\rank W_G = 1$, if and 
only if $\rank W_G(C) = 1 \text{ for all cliques } C \text{ in } G.$
\eee
If $G$ is a complete graph then $W_G$ specifies an $n\times n$ matrix
and the definitions of psd and rank-1 for the $G$-partial matrix $W_G$ coincide with
the regular definitions.

A cycle on $k$ nodes in graph $G$ is a $k$-tuple $(n_1, n_2, \ldots, n_k )$
such that $(n_1, n_2)$, $(n_2, n_3)$, $\ldots$, $(n_k, n_1)$ are edges in $G$. 
A cycle $(n_1, n_2, \ldots, n_k)$ in $G$ is minimal if no strict subset of 
$\{n_1, n_2, \ldots, n_k\}$ defines a cycle in $G$. In graph $G_1$ in Figure \ref{fig:graph1}
the 4-tuple $(1, 2, 3, 4)$ defines a minimal cycle. In graph $G_2$ in Figure \ref{fig:graph2} 
however the same 4-tuple is a cycle but not minimal. The minimal cycles in $G_2$ are 
$(1,2,3)$ and $(1, 3, 4)$. 
A graph is said to be \emph{chordal} if all its minimal cycles have at most 3 nodes. 
In Figure \ref{fig:graphs}, $G_2$ is a chordal graph while $G_1$ is not. 
A \emph{chordal extension} of a graph $G$ on $n$ nodes is a chordal graph $G_{ch}$ 
on the same $n$ nodes that contains $G$ as a subgraph. Note that all graphs have a chordal extension; the complete graph on the same set of vertices is a trivial chordal extension
of a graph. In Figure \ref{fig:graphs}, $G_2$ is a chordal extension of $G_1$.

Let $G_{ch}$ be any chordal extension of $G$.   
Define the following optimization problem over a Hermitian $G_{ch}$-partial matrix $W_{ch} := W_{G_{ch}}$, where the constraints \eqref{eq:opfW.1}-\eqref{eq:opfW.2} are imposed only on the index set $I_G \subseteq I_{G_{ch}}$, i.e., in terms of the $G$-partial submatrix $W_{ch}(G)$ of the $G_{ch}$-partial matrix $W_{ch}$.
\\
\textbf{Problem $\PP_{ch}$:}
\bqn
\underset{ W_{ch} }{\text{minimize}}  & & c(W_{ch}(G)) \\
\text{subject to} & & W_{ch}(G )\text{ satisfies } \eqref{eq:opfW.1}-\eqref{eq:opfW.2},\\
& & {W_{ch}} \succeq 0, \  \rank W_{ch} = 1.
\eqn
Let $\RR_{ch}$ be the rank-relaxation of $P_{ch}$.
\\
\textbf{Problem $\RR_{ch}$:}
\bqn
\underset{W_{ch}} {\text{minimize}}  & & c(W_{ch}(G)) \\
\text{subject to} & & W_{ch}(G) \text{ satisfies } \eqref{eq:opfW.1}-\eqref{eq:opfW.2}, 
\quad {W_{ch}} \succeq 0.
\eqn
Let $W_{ch}^*$ be an optimal solution of $\RR_{ch}$. If $W_{ch}^*$ is rank-1 then 
$W_{ch}^*$ also solves $\PP_{ch}$ optimally. Again, we say \emph{$\RR_{ch}$ is exact with respect to $\PP_{ch}$} if there exists an optimal solution $W_{ch}^*$ of $\RR_{ch}$ that has rank 1 and hence optimal for $\PP_{ch}$; see Remark \ref{rem:exactRel} for more details.

To illustrate, consider graph $G_1$ in Figure \ref{fig:graph1} and its chordal 
extension $G_2$ in Figure \ref{fig:graph2}. The cliques in $G_2$ are 
$\{ 1, 2\}$, $\{ 2, 3\}$, $\{ 3, 4\}$, $\{ 4, 1\}$, $\{ 1, 3\}$, $\{ 1, 2, 3\}$, $\{ 1, 3, 4\}$ and $\{ 4, 5\}$. 
Thus the constraint $W_{ch} \succeq 0 $ in $\RR_{ch}$ imposes positive semidefiniteness on 
$W_{ch}(C)$ for each clique $C$ in the above list. 
Indeed imposing  $W_{ch} (C) \succeq 0$ for maximal cliques $C$ of $G$ is sufficient, where a \emph{maximal clique} of a graph is a clique that is not a subgraph of another clique in the same graph.   This is because $W_{ch} (C) \succeq 0$ for a maximal clique $C$ implies
$W_{ch} (C') \succeq 0$ for any clique $C'$ that is a subgraph of $C$.
The maximal cliques in graph $G_2$ are $\{ 1, 2, 3\}$, $\{ 1, 3, 4\}$ and $\{ 4, 5\}$ and thus 
$W_{ch} \succeq 0$ is equivalent to $W_{ch}(C) \succeq 0$ for all maximal cliques $C$ listed above.
%
%
%
Even though listing all maximal cliques of a
general graph is NP-complete it can be done efficiently for a chordal graph.
This is because a graph is chordal if and only if it has a perfect elimination
ordering \cite{Fulkerson1965} and computing this ordering takes linear
time in the number of nodes and edges \cite{Rose1976}.  Given a perfect
elimination ordering  all maximal cliques $C$ can be enumerated and
$W_{ch}(C)$ constructed efficiently \cite{Fukuda99exploitingsparsity}.
Moreover the computation depends only on  network topology,
 not on operational data, and therefore can be done offline.
 For more details on chordal extension see \cite{Fukuda99exploitingsparsity}.
A special case of chordal relaxation is studied in \cite{Sojoudi2013} where
the underlying chordal extension extends every basis cycle of the network graph
into a clique.

\subsection{SOCP relaxation:  $\PP_2$ and $\RR_2$}

We say a $G$-partial matrix $W_G$ satisfies the \emph{cycle condition} if, over \emph{every} 
cycle $(n_1, \ldots, n_k)$ in $G$, we have
\begin{align}
\label{eq:cyclecond}
\angle [W_G]_{n_1  n_2} + \angle [W_G]_{n_2  n_3} + \ldots + \angle [W_G]_{n_k  n_1}  
= 0   \mod 2\pi.
\end{align}
\begin{remark} Consider any spanning tree of $G$. 
A ``basis cycle'' in $G$ is a cycle that has all but one of its edges common with
the spanning tree.  If \eqref{eq:cyclecond} holds over all basis cycles 
in $G$ with respect to a spanning tree then \eqref{eq:cyclecond} holds over 
all cycles of $G$ \cite{Biggs-1993-agt}.
\end{remark}

For any edge $e=(i, j)$ in $G$, $W_G(e)$ is the $2\times 2$ principal submatrix
of $W_G$ defined by the 2-clique $e$.
Define the following optimization problem over Hermitian $G$-partial matrices $W_G$.\\
\textbf{Problem $\PP_2$:}
\bqn
\underset{{W_G}}{\text{minimize}}  & & c(W_G) \\
\text{subject to} & & W_G \text{ satisfies } \eqref{eq:opfW.1}-\eqref{eq:opfW.2}
		\text{ and } \eqref{eq:cyclecond},
\\
& &{W_G}(e) \succeq 0, \ \rank {W_G}(e) = 1  \ \ \text{ for all } e \in E.
\eqn
Both the cycle condition \eqref{eq:cyclecond} and the rank-1 condition are nonconvex
constraints. Relaxing them, we get the following second-order cone program.\\
\textbf{Problem $\RR_2$:}
\bqn
\underset{{W_G}}{\text{minimize}}  & & c(W_G) \\
\text{subject to} & & W_G \text{ satisfies } \eqref{eq:opfW.1}-\eqref{eq:opfW.2},
\\
& &{W_G}(e) \succeq 0   \ \ \text{ for all } e \in E.
\eqn
For $e = (i,j)$ and Hermitian $W_G$ we have
\begin{align}
\label{eq:22psd}
W_G(e) \succeq 0 \quad \Leftrightarrow \quad [W_G]_{ii} [W_G]_{j j} \geq \left| [W_G]_{i j} \right|^2.
\end{align}
The right-hand side of \cref{eq:22psd} is a second-order cone constraint \cite{Boyd2004} and hence $\RR_2$ can be solved as an SOCP. If an optimal solution $W_G^*$ of $\RR_2$ is rank-1 and also satisfies the cycle condition then $W_G^*$ solves $\PP_2$ optimally and we say that relaxation \emph{$\RR_2$ is exact with respect to $\PP_2$}.



\subsection{Equivalent and exact relaxations}
\label{sec:mainRes}
So far, we have defined the problems $\PP_1$, $\PP_{ch}$ and $\PP_2$ and obtained 
their convex  relaxations $\RR_1$, $\RR_{ch}$ and $\RR_2$ respectively. 
We now characterize the relations among these problems. 

Let $p^*$ be the optimal cost of OPF. 
Let $p^*_1$, $p^*_{ch}$, $p^*_2$ be the optimal cost of 
$\PP_1$, $\PP_{ch}$, $\PP_2$ respectively and let $r^*_1$, $r^*_{ch}$, $r^*_2$ be 
the optimal cost of their relaxations $\RR_1$, $\RR_{ch}$, $\RR_2$ respectively. 
\begin{theorem}
\label{thm:main}
Let $G_{ch}$ denote any chordal extension  of $G$.  Then
\begin{enumerate}[(a)]
\item \label{thm.0} $p_1^* = p_{ch}^* = p_2^* = p^*$.
\item \label{thm.3} $r^*_1 = r^*_{ch} \geq r^*_2$.
	If $G$ is acyclic, then $r^*_1 = r^*_{ch} = r^*_2$.
\item \label{thm.1} $\RR_1$ is exact  iff  $\RR_{ch}$ is exact. $\RR_1$ and $\RR_{ch}$ are exact if 
	$\RR_{2}$ is exact.  If $G$ is acyclic, then
	$\RR_{2}$ is exact iff $\RR_1$ is exact.
\end{enumerate}
\end{theorem}
We make three remarks.  First, part (\ref{thm.0}) says that the optimal cost of 
$\PP_1$, $\PP_{ch}$ and $\PP_2$ are the same as that of OPF.   Our proof claims 
a stronger result: the underlying $G$-partial matrices in these problems are the
same.    Informally the feasible sets of these problems, and hence the problems
themselves, are equivalent and one can
construct a solution of OPF from a solution of any of these problems. 

Second, since $\PP_1$, $\PP_{ch}$ and $\PP_2$ are nonconvex we will solve
their relaxations $\RR_1$, $\RR_{ch}$ or $\RR_2$ instead.   Even though exactness is
defined to be a relation between each pair (e.g., $\RR_2$ is exact means 
$r_2^* = p_2^*$), part (\ref{thm.0}) says that if any pair is exact then the relaxed
problem is exact \emph{with respect to OPF} as well.
For instance if $\RR_2$ is exact with respect to $\PP_2$ then
any optimal $G$-partial matrix $W_G^*$ of $\RR_2$ 
satisfies \eqref{eq:cyclecond} and has rank $W_G^*(e) = 1$ for all $e\in E$.  
Our proof will construct a psd rank-1 $n \times n$ matrix $W^*$ from $W_G^*$ that 
is optimal for $\PP_1$.  The spectral decomposition of $W^*$ then
yields an optimal voltage vector $V^*$ in $\mathbb{V}$ for OPF.
Henceforth we will simply say that a relaxation $\RR_1/\RR_{ch}/\RR_2$ is ``exact''
instead of ``exact with respect to $\PP_1/\PP_{ch}/\PP_2$.''

Third, part (\ref{thm.1}) says that solving $\RR_1$ is the same as solving
$\RR_{ch}$ and, in the case where $G$ is acyclic (a \emph{tree}, since $G$ is assumed to be connected), is the same as solving $\RR_2$.
$\RR_1$ and $\RR_{ch}$ are SDPs while $\RR_2$ is an SOCP.
Though they can all be solved in polynomial time \cite{wolkowicz00, Boyd2004},
SOCP in general requires a much smaller computational effort than SDP.
Part (\ref{thm.3}) suggests that, when $G$ is a tree, we should always solve $\RR_2$.
When $G$ has cycles then there is a tradeoff between computational effort and
exactness in deciding between solving $\RR_2$ or $\RR_{ch}$/$\RR_1$.
As our simulation results in Section \ref{sec:numerics} confirm, 
if all maximal cliques of a chordal extension are available then solving $\RR_{ch}$ is 
\emph{always better} than solving $\RR_1$
as they have the same accuracy (in terms of exactness) but $\RR_{ch}$ is usually much
faster to solve for large sparse networks $G$.
Indeed $G$ is a subgraph of any chordal extension $G_{ch}$ of $G$ which is, in turn, a subgraph of the complete graph on $n$ nodes (denoted as $C_n$), and hence
$I_G \subseteq I_{G_{ch}} \subseteq I_{C_n}$.
Therefore, \emph{typically}, the number of variables is the smallest in $\RR_2$ $(|I_G|)$, 
the largest in $\RR_1$ $(|I_{C_n}|)$, with $\RR_{ch}$ in between.
However the actual number of variables in $\RR_{ch}$ is generally greater than
$|I_{G_{ch}}|$, depending on the choice of
the chordal extension $G_{ch}$.  Choosing a good $G_{ch}$ is nontrivial; see \cite{Fukuda99exploitingsparsity} for more details.
This choice however does not affect the optimal value $r_{ch}^*$.

\begin{corollary}
\label{coro:main}
\begin{enumerate}
\item If $G$ is acyclic then $p_* = p_1^* = p_{ch}^* = p_2^* \geq r_1^* = r_{ch}^* = r_2^{*}$.
\item If $G$ has cycles then $p_* = p_1^* = p_{ch}^* = p_2^* \geq r_1^* = r_{ch}^* \geq r_2^{*}$.
\end{enumerate}
\end{corollary}

Theorem \ref{thm:main} and Corollary \ref{coro:main} do not provide conditions that 
guarantee any of the relaxations $\RR_1, \RR_{ch}, \RR_2$ are exact.  
See \cite{Lavaei2012, Rantzer2011, Bose2011, Sojoudi2012PES, Zhang2013, LavaeiTseZhang2012, LamZhang2012, Sojoudi2013} for such sufficient conditions in the bus injection model. Corollary \ref{coro:main} implies that if $\RR_2$ is exact, so are $\RR_{ch}$ and
$\RR_1$.   Moreover Lemma \ref{lemma:setC} below relates the feasible sets of 
 $\RR_1, \RR_{ch}, \RR_2$, not just their optimal values.
It implies that $\RR_1, \RR_{ch}, \RR_2$ are
equivalent problems if $G$ has no cycles.  


\subsection{Proof of Theorem \ref{thm:main}}
We now prove that the feasible sets of OPF and $\PP_1, \PP_{ch}, \PP_2$ are equivalent
 when restricted to the underlying $G$-partial matrices.
Similarly, the feasible sets of their relaxations are equivalent when $G$ is a tree.
When any of the relaxations are exact we can construct an $n$-dimensional
 complex voltage vector $V \in \mathbb V$ that optimally solves OPF.

To define the set of $G$-partial matrices associated with $\PP_1, \PP_{ch}, \PP_2$
suppose $F$ is a graph on $n$ nodes such that $G$ is a subgraph of $F$, i.e., 
 $I_G \subseteq I_F$. 
An $F$-partial matrix $W_F$ is called an \emph{$F$-completion} of the $G$-partial 
 matrix $W_G$ if 
$$ [W_F]_{ij} = [W_G]_{ij} \text{ for all } (i, j) \in I_G \subseteq I_F,$$
i.e., $W_F$ agrees with $W_G$ on the index set $I_G$. 
If $F$ is $C_n$, the complete graph on $n$ nodes, then $W_F$ is an $n \times n$ matrix. $W_F$ is a Hermitian $F$-completion if $W_F = W_F^\herm$. 
$W_F$ is a psd $F$-completion if \emph{in addition} $W_F \succeq 0$. 
$W_F$ is a rank-1 $F$-completion if $\rank W_F =1$. It can be checked that if 
$W_G \not\succeq 0$ then $W_G$ does not have a psd $F$-completion.
  If $\rank W_G \neq 1$ then it does not have a rank-1 $F$-completion. Define
\begin{align*}
\FF_1 &:= \left\{ W_G \ \vert \ W_G\text{ satisfies } \eqref{eq:opfW.1}-\eqref{eq:opfW.2}, \right. \\
& \qquad \qquad\left.\exists \text{ psd rank-1  $C_n$-completion of } W_G  \right\}.
\end{align*}
Recall that for $W$, an $n \times n$ matrix, $W(G)$ is the $G$-partial matrix corresponding to the $I_G$ entries of $W$. Given an $n\times n$ psd rank-1 matrix $W$ that is feasible for  $\PP_1$, $W(G)$ is in $\FF_1$. 
Conversely given a $W_G \in \FF_1$, its psd rank-1 $C_n$-completion is a feasible 
solution for $\PP_1$. Hence $\FF_1$ is the set of $I_G$ entries of all $n \times n$ matrices 
feasible for $\PP_1$ and is nonconvex.  Define 
\begin{align*}
\FF_1^+ &:= \left\{ W_G \ \vert \ W_G\text{ satisfies } \eqref{eq:opfW.1}-\eqref{eq:opfW.2}, \right. \\
& \qquad \qquad\left.  \exists \text{ psd $C_n$-completion of } W_G  \right\}.
\end{align*}
$\FF_1^+$ is the set of $I_G$ entries of all $n \times n$ matrices feasible for $\RR_1$.
It is convex and contains $\FF_1$. 

Similarly define the corresponding sets for $\PP_{ch}$ and $\RR_{ch}$:
\begin{align*}
\FF_{ch} & := \left\{ W_G \ \vert  \ W_G\text{ satisfies } \eqref{eq:opfW.1}-\eqref{eq:opfW.2},\right. \\
& \qquad \qquad\left.  \exists \text{ psd rank-1  $G_{ch}$-completion of } W_G  \right\}, \\
\FF_{ch}^+ & := \left\{ W_G \ \vert \ W_G\text{ satisfies } \eqref{eq:opfW.1}-\eqref{eq:opfW.2}, \right. \\
& \qquad \qquad\left. \exists \text{ psd $G_{ch}$-completion of } W_G  \right\}.
\end{align*}
 $\FF_{ch}$  and $\FF_{ch}^+$ are the sets of $I_G$ entries of $G_{ch}$-partial matrices feasible for problems $\PP_{ch}$ and $\RR_{ch}$ respectively. 
 Again $\FF_{ch}^+$ is a convex set containing the nonconvex set $\FF_{ch}$. 
For problems $\PP_{2}$ and $\RR_{2}$ define:
\begin{align*}
\FF_2 &:= \left\{ W_G \ \vert \ W_G \text{ satisfies }\eqref{eq:opfW.1}-\eqref{eq:opfW.2} \text{ and } \eqref{eq:cyclecond},\right. \\
& \qquad \qquad  \left. {W_G}(e) \succeq 0, \ \rank {W_G}(e) = 1 \text{ for all  $e \in E$} \right\}, \\
\FF_2^+ &:= \left\{ W_G \ \vert  \ W_G \text{ satisfies }\eqref{eq:opfW.1}-\eqref{eq:opfW.2}, \right. \\
& \qquad \qquad\left. {W_G}(e) \succeq 0 \text{ for all  $e \in E$} \right\}.
\end{align*}
%
Informally the sets 
$\FF_1, \FF_1^+, \FF_{ch}, \FF_{ch}^+, \FF_2$ and $\FF_2^+$ describe
 the feasible sets of the various problems restricted to the $I_G$ entries of their respective 
 partial matrix variables. 

To relate the sets to the feasible set of OPF, consider the map $f$ from 
$\mathbb{C}^n$ to the set of $G$-partial matrices defined as: 
\begin{align*} f(V) := W_G \text{ where } [W_G]_{kk} &= |V_k|^2, \ k \in N, \text{and} \\ [W_G]_{j k} &= V_j V_k^H,  \ (j, k) \in E.
\end{align*}
Also, let $f(\mathbb V) := \{ f(V) \ \vert\ V\in \mathbb V \}$.

The sketch of the proof is as follows. We prove Theorem \ref{thm:main}(\ref{thm.0}) in Lemma \ref{lemma:setNC}
and then Theorem \ref{thm:main}(\ref{thm.3})  in Lemma \ref{lemma:setC} below.
Theorem \ref{thm:main}(\ref{thm.1}) then follows from these two lemmas.
\begin{lemma}
\label{lemma:setNC}
$f(\mathbb{V})  = \FF_1 = \FF_{ch} = \FF_2$.
\end{lemma}
\begin{IEEEproof}
First, we show that $f(\mathbb V) = \FF_1$. Consider $V \in \mathbb V$. Then $W = V V^H$ is feasible for $\PP_1$ and hence the $G$-partial matrix $W(G)$ is in $\FF_1$. Thus, $f(\mathbb V) \subseteq \FF_1$. To prove $\FF_1 \subseteq f(\mathbb V)$, consider the rank-1 psd $C_n$ completion of a $G$-partial matrix in $\FF_1$. Its unique spectral decomposition yields a vector $V$ that satisfies \eqref{eq:opf.1}--\eqref{eq:opf.2} and hence is in $\mathbb V$. Hence, $f(\mathbb V) = \FF_1$.

Now, fix a chordal extension $G_{ch}$ of $G$. We now prove:
\begin{align*}
\FF_{1} \ \subseteq \ \FF_{ch} \ \subseteq \ \FF_{2} \ \subseteq  \ \FF_1.
\end{align*}
To show $\FF_{1} \ \subseteq \ \FF_{ch}$, consider $W_G \in \FF_1$, and let $W$ be its
rank-1 psd $C_n$-completion. Then it is easy to check that $W(G_{ch})$ is feasible for $\PP_{ch}$ and hence $W_G$ is in $\FF_{ch}$ as well.

To show $\FF_{ch} \subseteq \FF_2$ consider a $W_G \in \FF_{ch}$ and 
its psd rank-1 $G_{ch}$-completion $W_{ch}$.  Since every edge $e$ of $G$
is a 2-clique in $G_{ch}$,  $W_G (e)=W_{ch}(e)$ is psd rank-1 by the definition
of psd and rank-1 for $W_{ch}$.
We are thus left to show that $W_G$ satisfies the cycle condition \eqref{eq:cyclecond}.
Consider the following statement $T_k$ for $3 \leq k \leq n$: \\
${S}_k$: For all cycles $(n_1, n_2, \ldots, n_k)$ of length $k$ in $G_{ch}$ we have:
\bqn
\angle {[W_{ch}]}_{n_1 n_2} + \angle {[W_{ch}]}_{n_2 n_3} + \ldots +
 \angle {[W_{ch}]}_{n_k n_1}  =  0   \mod 2\pi.
\eqn
For $k=3$, a cycle $(n_1, n_2, n_3)$  defines a 3-clique in $G_{ch}$ and thus 
$W_{ch}(n_1, n_2, n_3)$ is psd rank-1 and $W_{ch}(n_1, n_2, n_3) = u u^\herm$ for some 
$u := (u_1, u_2, u_3) \in \mathbb{C}^3$.
Then 
\begin{align*}
&\angle [W_{ch}]_{n_1 n_2} + \angle [W_{ch}]_{n_2 n_3} + \angle [W_{ch}]_{n_3 n_1}\\
&\quad = \angle \left[(u_1 u_2^H) (u_2 u_3^H)  (u_3 u_1^H) \right]  =  0   \mod 2\pi.
\end{align*}
Let $T_r$ be true for all $3 \leq r \leq k$ and consider a cycle 
$(n_1, n_2, \ldots, n_{k+1})$ of length $k+1$ in $G_{ch}$. 
Since $G_{ch}$ is chordal, this cycle must have a chord, i.e., an edge between two 
nodes, say, $n_1$ and $n_{k'}$, that are not adjacent on the cycle. 
Then $(n_1, n_2, \ldots, n_{k'})$ and $(n_1, n_{k'}, n_{k'+1}, \ldots, n_k)$ are two 
cycles in $G_{ch}$.  By hypothesis, $T_{k'}$ and $T_{k-k'+2}$ are true and hence
\begin{align*}
& \angle {[W_{ch}]}_{n_1 n_2} + \angle {[W_{ch}]}_{n_2 n_3} + \ldots + 
		\angle {[W_{ch}]}_{n_{k'} n_1} \\
& =\angle {[W_{ch}]}_{n_1 n_{k'}} + \angle {[W_{ch}]}_{n_{k'} n_{k' + 1}} + \ldots + 
		\angle {[W_{ch}]}_{n_{k} n_1} \\
& = 0 \mod 2 \pi.
\end{align*}
We conclude that $T_{k+1}$ is true by adding the above equations and using 
$\angle {[W_{ch}]}_{n_1 n_{k'}} = - \angle {[W_{ch}]}_{n_{k'}  n_{1}}\  \mod 2\pi$ since 
$W_{ch}$ is Hermitian. By induction, $W_{ch}$ satisfies the cycle condition. Also, $W_G = W_{ch}(G)$ satisfies the cycle condition and hence in $\FF_2$. This completes the proof of $\FF_{ch}  \subseteq \FF_2$.

To show $\FF_2 \subseteq \FF_1$ suppose $W_G \in \FF_2$. 
We now construct a psd rank-1 $C_n$-completion of $W_G$ to show $W_G \in \FF_1$.
Define $\theta \in \mathbb{C}^n$ as follows. Let $\theta_1 := 0$. 
For $j \in N\setminus \{1\}$ let $(1, n_2)$, $(n_2, n_3)$, $\ldots, (n_k, j)$ be 
any path from node 1 to node $j$. Define
\bqn
\theta_j := - (\angle [W_G]_{1 n_2} + \angle [W_G]_{n_2 n_3} + \ldots + \angle [W_G]_{n_k j})
\mod  2\pi.
\eqn
Note that the above definition is well-defined:
 if there is another sequence of edges from node 1 to node $j$, the above relation 
 still defines $\theta_j$ uniquely  because $W_G$ satisfies the cycle condition.
 Let
 \bqn
 V  & := &  \left[ \sqrt{[W_G]_{11} }\, e^{\ii \theta_1}, \ \ \cdots \ \ 
 		\sqrt{[W_G]_{nn} }\, e^{\ii \theta_n} \right].
\eqn
Then it can be verified that $W := VV^H$ is a psd rank-1 $C_n$-completion of $W_G$.
Hence $W_G \in \mathbb \FF_1$. This completes the proof of the lemma.
\end{IEEEproof}

%
\begin{lemma}
\label{lemma:setC}
$\FF_1^+ = \FF_{ch}^+  \subseteq   \FF_2^+$. 
If $G$ is acyclic, then $\FF_1^+ = \FF_{ch}^+  =  \FF_2^+$.
\end{lemma}

\begin{IEEEproof}
It suffices to prove
\begin{align}
\label{eq:setInc.2}
\FF_{ch}^+ \ \subseteq \ \FF_{1}^+ \ \subseteq \ \FF_{ch}^+ \ \subseteq \ \FF_{2}^+.
\end{align}
To show $\FF_{ch}^+ \subseteq \FF_1^+$, suppose $W_G \in \FF_{ch}^+$.
Let $W_{ch}$ be a psd $G_{ch}$-completion of $W_G$ for a chordal extension
 $G_{ch}$.
Since any psd partial matrix  on a chordal graph has a psd $C_n$-completion 
\cite[Theorem 7]{Grone1984}, $W_{ch}$ has a psd $C_n$-completion.
Obviously, any psd $C_n$-completion of $W_{ch}$ is also a psd $C_n$-completion 
of $W_G$, i.e., $W_G\in \FF_1^+$.
The relation $\FF_{1}^+ \ \subseteq \ \FF_{ch}^+ \ \subseteq \ \FF_{2}^+$ follows 
a similar argument to the proof of Lemma \ref{lemma:setNC}. 

If $G$ is acyclic, then $G$ is itself chordal and hence $W_G$ has a psd
$C_n$-completion, i.e., $\FF_2^+ \subseteq \FF_1^+$.  
This implies $\FF_1^+ = \FF_{ch}^+  =  \FF_2^+$.
\end{IEEEproof}

To prove Theorem \ref{thm:main}(\ref{thm.1}) note that parts (\ref{thm.0}) and
(\ref{thm.3}) imply
\bqn
p^* = p_1^* = p_{ch}^* = p_2^* & \geq & r_1^* = r_{ch}^* \geq r_2^*.
\eqn
Hence $\RR_1$ is exact $(p_1^* = r_1^*)$ iff $\RR_{ch}$ 
is exact $(p_{ch}^* = r_{ch}^*$).  If $\RR_2$ is exact, i.e., $p_2^* = r_2^*$, then both inequalities
above become equalities, proving Theorem \ref{thm:main}(\ref{thm.1}).
This completes the proof of Theorem \ref{thm:main}.



%% file: BranchFlow2.tex

\section{Branch flow model and SOCP relaxation}
\label{sec:bfm}

\subsection{OPF formulation}

The branch flow model of \cite{Farivar-2013-BFM-TPS}  adopts
a directed connected graph $\tilde G = (N, \tilde E)$ to represent
a power network where each node in $N := \{1, \dots, n\}$ represents a bus and each edge in $\tilde E$ represents a line. The orientations of the edges are taken to be arbitrary. Denote the directed edge from bus $i$ to bus $j$ by $i \rightarrow j \in \tilde{E}$ and define $m := |\tilde E|$ as  the number of directed edges in $G$. For each edge $i \rightarrow j \in \tilde E$, define the following quantities:
\begin{itemize}
\item $z_{ij}$: The complex impedance on the line. Thus $z_{ij}= 1/y_{ij}$.
\item $I_{ij}$: The complex current from bus $i$ to bus $j$.
\item $S_{ij}$: The {\em sending-end} complex power from buses $i$ to $j$.
\end{itemize}
Recall that for each node $i \in N$, $V_i$ is the complex voltage at bus $i$ and $s_i$ is the net complex power injection (generation minus load) at bus $i$.

The \emph{branch flow model} of \cite{Farivar-2013-BFM-TPS}
 is defined by the following set of power flow equations:
\begin{subequations}
\label{eq:bfm}
\begin{align}
\label{eq:bfm.1}
s_j  = \sum_{k: j\rightarrow k} S_{jk} - \sum_{i: i\rightarrow j} \left( S_{ij} - z_{ij} |I_{ij}|^2 \right) 
\; \text{for } j \in N, \\
\label{eq:bfm.2}
S_{ij}  = V_i \ I_{ij}^\herm \; \text{and} \; I_{ij}  = y_{ij} (V_i - V_j)
\quad \text{for } i \to j \in \tilde{E},
\end{align}
\end{subequations}
where \eqref{eq:bfm.1} imposes power balance at each bus and \eqref{eq:bfm.2} defines branch power and describes Ohm's law. The power injections at all buses  satisfy
\bq
 \underline{s}_j  \leq s_j \leq \overline{s}_j  \ \ \ \ \text{ for } j \in N, 
 \label{eq:opfB.2a}
\eq
where $\underline{s}_j$ and $\overline{s}_j$ are known limits on the net generation
at bus $j$. 
It is often assumed that the slack bus (node 1) has a generator and there is no
limit of $s_1$; in this case $-\underline{s}_j = \overline{s}_j = \infty$.
As in the bus injection model, we can eliminate the variables $s_j$ by combining
\eqref{eq:bfm.1} and \eqref{eq:opfB.2a} into:
\bq
\underline{s}_j  \ \leq \,
 \sum_{k: j\rightarrow k} S_{jk} - \sum_{i: i\rightarrow j} \left( S_{ij} - z_{ij} |I_{ij}|^2 \right) 
 \leq  \overline{s}_j \ \ \text{for } j \in N.
\label{eq:bfm.1a}
\eq
All voltage magnitudes are  constrained as follows:
 \bq
\underline{V}_j \leq |V_j| \leq \overline{V}_j  \ \  \text{ for } j \in N,
  \label{eq:opfB.2}
\eq
where $\underline{V}_j$ and $\overline{V}_j$ are known lower and upper voltage limits,
with  $|V_1| = 1 = \underline{V}_1 = \overline{V}_1$.
Denote the variables in the branch flow model by $\tilde{x} := (S, I, V) \in \mathbb{C}^{n+2m}$.  
These constraints define  the feasible set of the OPF  problem in the branch flow model:
\begin{align}
\mathbb{X} := \{ \tilde{x} \in \mathbb{C}^{n+2m} \ \vert \ \tilde{x} \text{ satisfies } 
				\eqref{eq:bfm.2}, \eqref{eq:bfm.1a}, \eqref{eq:opfB.2} \}.
\label{eq:bfmFS}
\end{align}

To define OPF, consider a cost function $c(\tilde x)$. For example, if the objective is to minimize the real power loss in the network, then we have
$$c(\tilde x) = \sum_{j \in N} \Re s_j = \sum_{j \in N} \text{Re} \left[\sum_{k: j\rightarrow k} S_{jk} - \sum_{i: i\rightarrow j} \left( S_{ij} - z_{ij} |I_{ij}|^2 \right) \right].$$
Similarly, if the objective is to minimize the weighted sum of real power generation in the network, then
\begin{align*}
c(\tilde x) &= \sum_{j \in N} c_j \left(\Re s_j - p_j^{d} \right)\\
&= \sum_{j \in N} c_j  \left[ \text{Re} \left(\sum_{k: j\rightarrow k} S_{jk} - \sum_{i: i\rightarrow j} \left( S_{ij} - z_{ij} |I_{ij}|^2 \right) \right) - p_j^d \right],
\end{align*}
where $p_j^{d}$ is the given real power demand at bus $j \in N$.

\noindent
\textbf{Optimal power flow problem $OPF$:}
\begin{align}
\label{eq:bfmOPF}
& \underset{\tilde x } {\text{minimize}}   \ \  c(\tilde x) \quad
\text{ subject to } \   \tilde x \in \XX.
\end{align}
Since \eqref{eq:bfm} is quadratic, $\mathbb{X}$ is generally a nonconvex set.
As before, OPF is a nonconvex problem.

\subsection{SOCP relaxation: $\tilde{\PP}_2$, $\tilde{\RR}_2^{nc}$ 
		and $\tilde{\RR}_2$}

The  SOCP relaxation of \eqref{eq:bfmOPF} developed in \cite{Farivar-2013-BFM-TPS}
consists of two steps.  First, we use \eqref{eq:bfm.2} to eliminate the phase angles from 
the complex voltages $V$ and currents $I$ to obtain for each $i \rightarrow j \in \tilde E$,
\bq
v_j & = & v_i - 2 \text{ Re }(z_{ij}^H S_{ij}) + |z_{ij}|^2 \ell_{ij},
\label{eq:Kirchhoff.2c}
\\
\ell_{ij} v_i & = & |S_{ij}|^2.
\label{eq:Kirchhoff.2d}
\eq
where $v_i := |V_i|^2$ and $\ell_{ij} := |I_{ij}|^2$.  
This is the model first proposed by Baran-Wu in
\cite{Baran1989a, Baran1989b} for distribution systems.
Second the  quadratic equalities in \eqref{eq:Kirchhoff.2d} are nonconvex;
relax them  to inequalities:
\bq
\ell_{ij} v_i & \geq &  |S_{ij}|^2  \ \ \ \ \text{ for }   i \rightarrow j\in \tilde{E}.
\label{eq:Kirchhoff.2e}
\eq

Let $x := (S, \ell, v) \in \mathbb R^{n+3m}$ denote the new variables. Note that we use $S$ to denote both a complex variable in $\mathbb C^{m}$ and the real variables $(\Re S, \Im S)$ in $\mathbb R^{2m}$ depending on context.
Define the nonconvex set:
 \bqn
{\XX}_2^{nc} & := & \{ x \in \mathbb R^{n+3m} \ \vert \ x \text{ satisfies } 
		(\ref{eq:bfm.1a}), \eqref{eq:opfB.2}, \eqref{eq:Kirchhoff.2c}, \eqref{eq:Kirchhoff.2d} \},
\eqn
and the convex superset that is a second-order cone:
\bqn
{\XX}_2^+ & := & \{ x \in \mathbb R^{n+3m} \ \vert \ x \text{ satisfies } 
		(\ref{eq:bfm.1a}), \eqref{eq:opfB.2}, \eqref{eq:Kirchhoff.2c}, \eqref{eq:Kirchhoff.2e} \}.
\eqn
As we discuss below solving OPF over $\XX_2^+$ is an SOCP and hence
efficiently computable. Whether the solution of the SOCP relaxation yields an optimal for OPF depends on two factors \cite{Farivar-2013-BFM-TPS}: 
(a) whether the optimal solution over $\XX_2^+$ actually lies in $\XX_2^{nc}$, (b) whether the phase angles of $V$ and $I$ can be recovered from such a solution, as we now explain.

For an $n \times 1$ vector $\theta \in [-\pi, \pi)^n$  define the map
 $h_\theta: \mathbb{R}^{n+3m} \rightarrow \mathbb{C}^{n+2m}$ by
$h_\theta(S, \ell, v) = (S, I, V)$ where  
\bqn
V_i & := & \sqrt{v_i} \ e^{\ii \theta_i}  \quad \text{for } i \in N,
\\
I_{ij} & := & \sqrt{\ell_{ij}}\ e^{\ii (\theta_i - \angle S_{ij})} \quad \text{for } i \rightarrow j \in \tilde E.
\eqn
Given an $ x := (S, \ell, v) \in \XX_2^+$ our goal is to find $\theta$
so that $h_\theta(x) \in \XX$ is feasible for OPF. To determine whether such a $\theta$ exists, define $\beta({x}) \in \mathbb R^m$ by
\bq
\beta_{ij}( x) & := & \angle \left( v_i - z_{ij}^H S_{ij} \right) \quad\text{for } i \rightarrow j\in \tilde E.
\label{eq:defb.1}
\eq
Essentially, $x\in \XX_2^+$ implies a phase angle difference across each line $i \rightarrow j \in \tilde E$ given by $\beta_{ij}(x)$ \cite[Theorem 2]{Farivar-2013-BFM-TPS}.  
We are interested in the set of $x$ such that $\beta_{ij}(x)$ can be expressed
as $\theta_i - \theta_j$ where $\theta_i$ can be the phase of voltage at node $i \in N$.
In particular, let $C$ be the $n\times m$ incidence matrix of $\tilde{G}$ defined as
\bqn
C_{ie} & = & \begin{cases}
		1 & \text{ if edge $e \in \tilde E$ leaves node $i \in N$}, \\
		-1 & \text{ if edge $e \in \tilde E$ enters node $i \in N$}, \\
		0 & \text{ otherwise}.
		\end{cases}
		\quad\quad  
\eqn
The first row of $C$ corresponds to the slack bus.
Define the $m \times (n-1)$ {\em reduced} incidence matrix $B$
obtained from $C$ by removing the first row and taking the transpose.
%
Consider the set of $x$ such that 
\bq
\exists \  \theta \text{ that solves }\ \ B\theta  = \beta(x)  \mod 2\pi.
\label{eq:cyclecond.2}
\eq
A solution $\theta$, if exists, is unique in  $[-\pi, \pi)^n$.
Moreover the necessary and sufficient condition for the existence of a solution to
 \eqref{eq:cyclecond.2} has a familiar interpretation:
the implied voltage angle differences $\beta(x)$ sum to zero (mod $2\pi$) around any cycle
 \cite[Theorem 2]{Farivar-2013-BFM-TPS}.

Define the set:
 \bqn
{\XX}_2  :=  \{ x \in \mathbb R^{n+3m} \ \vert \ x \text{ satisfies } 
		(\ref{eq:bfm.1a}), \eqref{eq:opfB.2}, \eqref{eq:Kirchhoff.2c}, \eqref{eq:Kirchhoff.2d},
		\eqref{eq:cyclecond.2}  \}.
\eqn
Clearly $\XX_2 \subseteq \XX_2^{nc} \subseteq X_2^+$.
These three sets define the following optimization problems.\footnote{Recall that cost $c(\cdot)$ was defined over $(S, I, V) \in \mathbb C^{n+2m}$. For the cost functions considered, it can be equivalently written as a function of $(S, \ell, v) \in \mathbb R^{n+3m}$.}
\\
\noindent
\textbf{Problem $\tilde{\PP}_2$:}
\begin{subequations}
\begin{align*}
& \underset{ x } {\text{minimize}}   \ \  c( x) \quad
 \text{ subject to } \quad  x \in \XX_2.
\end{align*}
\end{subequations}

\noindent
\textbf{Problem $\tilde{\RR}_2^{nc}$:}
\begin{subequations}
\begin{align*}
& \underset{ x} {\text{minimize}}   \ \  c( x) \quad
\text{ subject to} \quad  x \in \XX_2^{nc}.
\end{align*}
\end{subequations}

\noindent
\textbf{Problem $\tilde{\RR}_2$:}
\begin{subequations}
\begin{align*}
& \underset{ x } {\text{minimize}}   \ \  c( x) \quad
\text{ subject to } \quad  x \in \XX_2^+.
\end{align*}
\end{subequations}
We say \emph{$\tilde \RR_2$ is exact with respect to $\tilde \RR_2^{nc}$} if there exists an optimal solution $x^*$ of $\tilde{\RR}_2$ that attains equality in \eqref{eq:Kirchhoff.2e}, 
i.e., $x^*$  lies in $\XX_2^{nc}$. 
We say \emph {$\tilde \RR_2^{nc}$ is exact with respect to $\tilde \PP_2$} 
if there exists an optimal solution $x^*$ of $\tilde \RR_2^{nc}$ that satisfies
\eqref{eq:cyclecond.2}, i.e., $x^*$ lies in $\mathbb X_2$ and solves $\tilde \PP_2$ optimally. 

The problems $\tilde \PP_2$ and $\tilde \RR_2^{nc}$ are nonconvex and hence NP-hard,
but problem $\tilde \RR_2$ is an SOCP and hence can be solved in polynomial time \cite{tsuchiya1997polynomial, Boyd2004}. 
%
%
Let $p^*$ be the optimal cost of OPF \eqref{eq:bfmOPF} in the branch flow model. Let $\tilde p_2^*$, $\tilde r_2^{nc}$, $\tilde r_2^*$ be the optimal costs of 
$\tilde{\PP}_2$, $\tilde{\RR}_2^{nc}$, $\tilde{\RR}_2$ respectively. 
The next result follows directly from \cite[Theorems 2, 4]{Farivar-2013-BFM-TPS}.

\begin{theorem}
\label{thm:bfmMain}
\bee[(a)]
\item There is a bijection between $\XX$ and $\XX_2$.
\item  $p^* = \tilde p_2^* \geq \tilde r_2^{nc} \geq \tilde r_2^*$ where the first inequality is an equality 
	 if $\tilde G$ is acyclic.
\eee
\end{theorem}

We make two remarks on this relaxation over radial (tree) networks $\tilde G$. 
First, for such a graph, Theorem \ref{thm:bfmMain} says that if $\tilde \RR_2$ is exact with respect to 
$\tilde{\RR}_2^{nc}$, then it is exact with respect to OPF \eqref{eq:bfmOPF}. Indeed, for any optimal solution $x^*$ of $\tilde{\RR}_2$ that attains equality in \eqref{eq:Kirchhoff.2e}, the relation in \eqref{eq:cyclecond.2} always has a unique solution $\theta^*$ in $ [-\pi, \pi)^n$ and hence $h_{\theta^*}(x^*)$ is optimal for OPF. 

Second, Theorem \ref{thm:bfmMain} does not provide conditions that guarantee $\tilde{\RR}_2$ or $\tilde{\RR}_2^{nc}$ is exact.   See 
\cite{Farivar2011-VAR-SGC, Farivar-2013-BFM-TPS, Gan-2012-BFMt, Li-2012-BFMt} for sufficient conditions for exact SOCP relaxation in radial networks.
Even though, here, we define a relaxation to be exact as long as one of its optimal
solutions satisfies the constraints of the original problem, all the sufficient conditions
in these papers guarantee that \emph{every} optimal solution of the relaxation is optimal 
for the original problem.

%% file: modelEquiv2.tex
\section{Equivalence of bus injection and branch flow models}
\label{sec:eq}

In this section we establish equivalence relations between the
bus injection model and the branch flow model and their relaxations. 
Specifically we establish two sets of bijections (a) between the feasible sets of problems $\PP_2$ and $\tilde \PP_2$, i.e., $\FF_2$ and $\XX_2$, 
and (b) between the feasible sets of problems 
$\RR_2$ and $\tilde \RR_2$, i.e., $\FF_2^+$ and $\XX_2^+$.

For a Hermitian $G$-partial matrix $W_G$, define the $(n+3m) \times 1$ vector $x=(S, \ell, v) := g(W_G)$ as follows. For $i\in N$ and $i \rightarrow j \in \tilde E$,
\begin{align}
v_i & := [W_G]_{ii},
\label{eq:g.1}
\\
S_{ij} &:= y_{ij}^\herm \left( [W_G]_{ii} - [W_G]_{ij} \right),
\label{eq:g.2}
\\
\ell_{ij} & := |y_{ij}|^2 \left( [W_G]_{ii} + [W_G]_{jj} - [W_G]_{ij} - [W_G]_{ji} \right).
\label{eq:g.3}
\end{align}
Define the mapping $g^{-1}$ from  $\mathbb R^{n+3m}$ to the set of Hermitian $G$-partial matrices as follows. Let $W_G := g^{-1}(x)$ where
\begin{align}
[W_G]_{ii} & :=  \, v_i  \quad \text{for } i\in N, 
\label{eq:ginv.1} \\
[W_G]_{ij} & :=  \,  v_i - z_{ij}^\herm S_{ij}  \ = \ [W_G]_{ji}^\herm \quad \text{for } i \to j \in \tilde E. 
\label{eq:ginv.2}
\end{align}

The next result implies that $g$ and $g^{-1}$ restricted to $\FF_2^+$ $(\FF_2)$
and $\XX_2^+$ $(\XX_2)$ respectively are indeed inverse of each other.
This establishes a bijection between the respective sets.
\begin{theorem}
\label{thm:branch}
\bee[(a)]
\item The mapping $g: \FF_2 \rightarrow \XX_2$ is a bijection with $g^{-1}$ as its inverse.

\item The mapping $g: \FF_2^+ \rightarrow \XX_2^+$ is a bijection with $g^{-1}$ as its inverse.
\eee
\end{theorem}

Before we present its proof we make three remarks.
First, Lemma \ref{lemma:setNC} implies a bijection between  $\FF_2$ and
the feasible set $V$ of OPF in the bus injection model.
Theorem \ref{thm:bfmMain}(a) implies a bijection between $\XX_2$ and
the feasible set $\XX$ of OPF in the branch flow model. 
Theorem \ref{thm:branch} hence implies a bijection between the
feasible sets $\mathbb V$ and $\XX$ of OPF in the bus injection model 
and the branch flow model respectively. It is in this sense that these two 
models are equivalent.

Second, it is important that we utilize both models because some relaxations are
much easier to formulate and some sufficient conditions for exact relaxation are 
much easier to prove in one model than the other.  
For instance the semidefinite relaxation of power flows has a much cleaner 
formulation in the bus injection model.  
The branch flow model especially for radial networks has a convenient 
recursive structure that not only allows a more efficient computation of power flows
e.g. \cite{Kersting2002, Shirmohammadi1988, ChiangBaran1990}, but
also plays a crucial role in proving the sufficient conditions for exact relaxation 
in \cite{Gan-2013-BFMt-CDC, Gan-2013-BFMt-TAC}.
Since the variables in
the branch flow model correspond directly to physical quantities such as
branch power flows and injections it is sometimes more convenient in
applications.

Third, define the set of $G$-partial matrices that are in $\FF_2^+$
but do not satisfy the cycle condition \eqref{eq:cyclecond}:
\begin{align}
\FF_2^{nc} &:= \left\{ W_G \ \vert \  W_G \text{ satisfies }
	\eqref{eq:opfW.1}-\eqref{eq:opfW.2}, \right.\notag\\
& \qquad \left.W_G(e) \succeq 0, \rank {W_G}(e) = 1 \text{ for  $e \in E$} \right\}.\label{eq:W2nc}
\end{align}
Clearly, $\FF_2 \subseteq \FF_2^{nc} \subseteq \FF_2^+$.
Then the same argument as in Theorem \ref{thm:branch} implies that $g$ and 
$g^{-1}$ define a bijection between $\FF_2^{nc}$ and $\XX_2^{nc}$.

\begin{figure*}[!bht]
\centering
\includegraphics[width=0.88\textwidth]{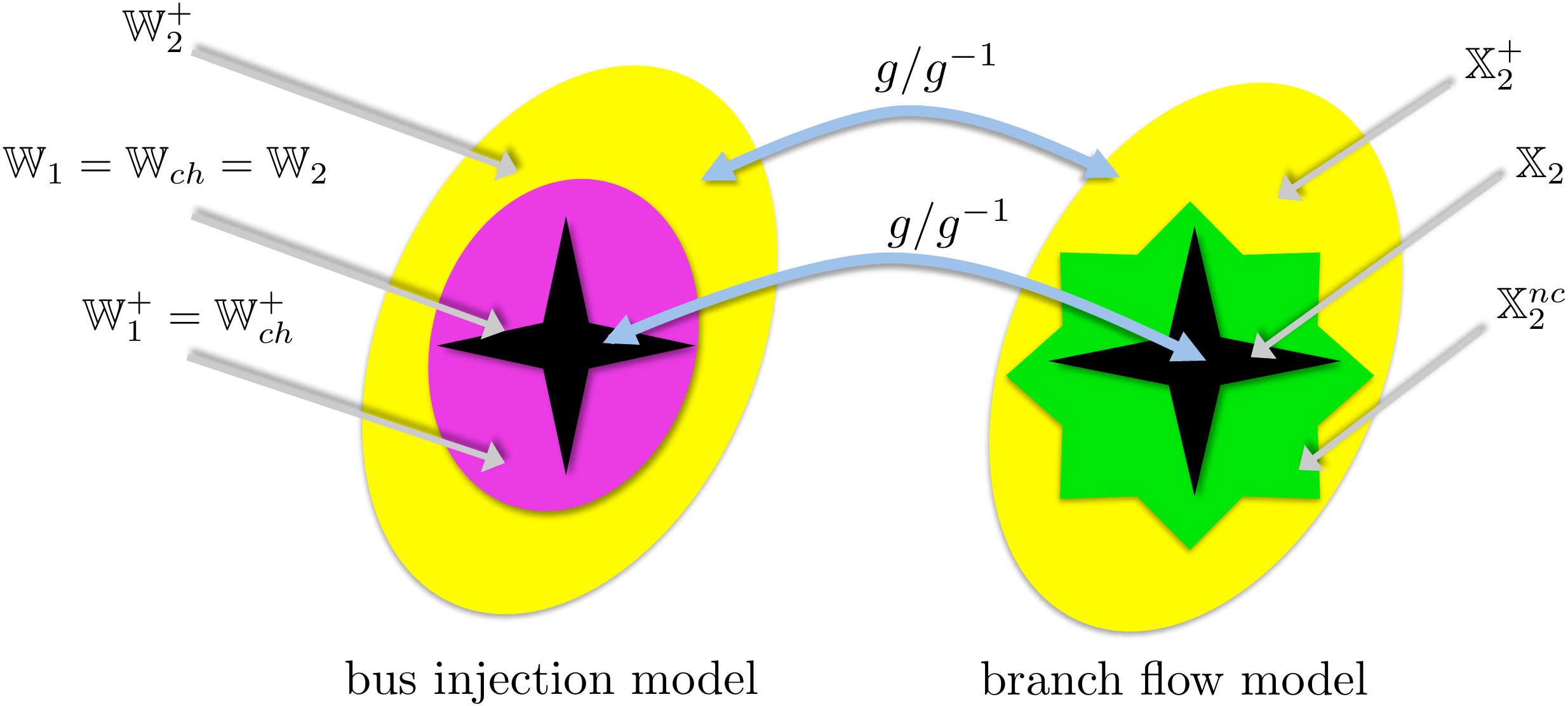}
\caption{Feasible sets of conic formulations and their relaxations, and the relations among
these sets. The sets $\mathbb W_1 = \mathbb W_{ch} = \mathbb W_2$ on the left are the nonconvex feasible sets
of equivalent OPF problems $\PP_1$, $\PP_{ch}$, $\PP_2$ respectively in the bus injection
model, and $\mathbb W_1^+ = \mathbb W_{ch}^+ \subseteq \mathbb W_2^+$ are 
the convex feasible sets of their respective relaxations $\RR_1, \RR_{ch}$, $\RR_2$.
On the right, $\mathbb X_2$ is the nonconvex feasible set of an equivalent OPF problem $\tilde \PP_2$
in the branch flow model. $\mathbb X_2^{nc}$ is the nonconex feasible set of the relaxation
$\tilde \RR_2^{nc}$ obtained by eliminating the voltage phase angles and $\XX_2^+$ is
the convex feasible set of the relaxation $\tilde \RR_2$.
The equivalence of the sets $\FF_2$ (or $\FF_2^+$) and $\XX_2$ (or $\XX_2^+$)
is represented by the linear maps $g/ g^{-1}$. When $G$ is a tree, $\FF_1^+ = \FF_{ch}^+ = \FF_2^+$ in the bus injection model and $\XX_2^{nc} = \XX_2^+$ in the branch flow model.
Note that neither of $\mathbb W_1^+$ and $\mathbb X_2^{nc}$ (or, more precisely $g^{-1}(\mathbb X_2^{nc})$ ) contains the other.
}
\label{fig:setRelations}
\end{figure*}

\begin{IEEEproof}[Proof of Theorem \ref{thm:branch}]
We only prove part (a);  part (b) follows similarly. 
Recall the definitions of sets $\FF_2$ and $\XX_2$:
\begin{align*}
\FF_2 & :=  \left\{ W_G \ \vert \ W_G \text{ satisfies }\eqref{eq:opfW.1}-\eqref{eq:opfW.2} \text{ and } \eqref{eq:cyclecond},\right. \\
& \qquad \qquad \left. {W_G}(e) \succeq 0, \ \rank {W_G}(e) = 1 \text{ for all  $e \in E$} \right\},\\
{\XX}_2  &:=  \{ x \in \mathbb R^{n+3m} \ \vert \ x \text{ satisfies } 
		(\ref{eq:bfm.1a}), \eqref{eq:opfB.2}, \eqref{eq:Kirchhoff.2c}, \eqref{eq:Kirchhoff.2d},
		\eqref{eq:cyclecond.2}  \}.
\end{align*}
We need to show that
\bee
\item[(i)] $g(\FF_2) \subseteq \XX_2$ so that $g: \FF_2 \rightarrow \XX_2$ is well defined.
\item[(ii)] $g$ is injective, i.e., $g(x) \neq g(x')$ if $x\neq x'$.
\item [(iii)] $g$ is surjective and hence its inverse exists; moreover $g^{-1}$ 
	defined in \eqref{eq:ginv.1}--\eqref{eq:ginv.2} is indeed $g$'s inverse.
\eee
The proof of (i) is similar to that of (iii) and omitted.  That $g$ is injective follows directly 
from \eqref{eq:g.1}--\eqref{eq:g.3}.   To prove (iii), we need to show that
given any $x := (S, \ell, v) \in \XX_2$, $W_G := g^{-1}(x)$ defined by 
\eqref{eq:ginv.1}--\eqref{eq:ginv.2} is in $\FF_2$ and $x = g(W_G)$.   
We now prove this in four steps.

\noindent
{\em Step 1: Proof that $W_G$ satisfies \eqref{eq:opfW.1}--\eqref{eq:opfW.2}.}
Clearly \eqref{eq:opfW.2} follows from \eqref{eq:opfB.2}.   We now show that
\eqref{eq:opfW.1} is equivalent to \eqref{eq:bfm.1a}.   For node $j \in N$, separate the edges in the
summation in \eqref{eq:opfW.1} into outgoing edges $j \rightarrow k \in \tilde{E}$
from node $j$  and incoming edges $k\rightarrow j\in \tilde E$ to node $j$.   
For each  incoming edge $k\rightarrow j\in \tilde E$ we have from 
\eqref{eq:ginv.1}--\eqref{eq:ginv.2}
\bqn
 [W_G]_{jj} - [W_G]_{jk} & =  v_j - \left( v_k - z_{kj}^H S_{kj} \right)^H 
 \\
 & =   - \left( v_k - v_j - z_{kj} S_{kj}^H \right)
 \\
 & =  - \left( z_{kj}^H S_{kj} - |z_{kj}|^2 \ell_{kj} \right),
\eqn
where the last equality follows from \eqref{eq:Kirchhoff.2c}.
Substituting this and \eqref{eq:ginv.1}--\eqref{eq:ginv.2} into \eqref{eq:opfW.1} we get,
for each $j\in N$:
\begin{align*}
&   \sum_{k: (j,k)\in E} \left( [W_G]_{jj} - [W_G]_{jk} \right) y_{jk}^H
\\
& \qquad =  \sum_{k: j\rightarrow k \in \tilde E} \left( [W_G]_{jj} - [W_G]_{jk} \right) y_{jk}^H \\
     & \qquad \qquad + \sum_{k: k\rightarrow j \in \tilde E} \left( [W_G]_{jj} - [W_G]_{jk} \right) y_{jk}^H
\\
& \qquad =  \sum_{k: j\rightarrow k \in \tilde E} \left( v_j - (v_j - z_{jk}^\herm S_{jk} )  \right) y_{jk}^H \\
       & \qquad \qquad - \sum_{k: k\rightarrow j \in \tilde E} \left( z_{kj}^H S_{kj} - |z_{kj}|^2 \ell_{kj} \right) y_{kj}^H
\\
& \qquad =  \sum_{k: j\rightarrow k} S_{jk} 	
	 \ - \sum_{k: k\rightarrow j} \left( S_{kj} - z_{kj} \ell_{kj} \right). 
\end{align*}
Hence, \eqref{eq:opfW.1} is equivalent to \eqref{eq:bfm.1a}.

\noindent
{\em Step 2: Proof that $W_G$ satisfies \eqref{eq:cyclecond}.}   
Without loss of generality let $c := (1, 2, \dots, k)$ be a cycle.  
For each {\em directed} edge $i\rightarrow j \in \tilde E$, recall 
$\beta_{ij}(x) :=  \angle (v_i - z_{ij}^H S_{ij})$ defined in \eqref{eq:defb.1} 
and define $\beta_{ji}(x) = - \beta_{ij}(x)$ in the opposite direction.
Since $x = (S, \ell, v)$ satisfies \eqref{eq:cyclecond.2},
\cite[Theorem 2]{Farivar-2013-BFM-TPS} implies that
\bq
\beta_{12}(x) + \dots + \beta_{k1}(x) & = & 0  \mod 2\pi,
\label{eq:sumb}
\eq
where each $(i,j)$ in $c$ may be in the same  or opposite
orientation as the orientation of the directed graph $\tilde G$.
Observe from \eqref{eq:ginv.2} that, for each directed edge $i\rightarrow j\in \tilde E$,
$\angle [W_G]_{ij} = \beta_{ij}(x)$ and $\angle [W_G]_{ji} = \beta_{ji}(x)$.  
Hence \eqref{eq:sumb} is equivalent to \eqref{eq:cyclecond}, 
i.e., $\sum_{(i,j)\in c} \angle [W_G]_{ij} = 0 \mod 2\pi$.

\noindent
{\em Step 3: Proof that $W_G(e) \succeq 0,  \rank {W_G}(e) = 1$  for all  $e \in E$.}
For each edge $i \rightarrow j \in \tilde E$ we have
\begin{align}
& [W_G]_{ii} [W_G]_{jj} - [W_G]_{ij} [W_G]_{ij}^H \\
& \quad =  v_i v_j - \left| v_i - z_{ij}^H S_{ij} \right|^2
\notag
\\
& \quad = v_i v_j - \left( v_i^2 - v_i (z_{ij}S_{ij}^H + z_{ij}^H S_{ij}) + |z_{ij}|^2 |S_{ij}|^2 \right)
\notag
\\
& \quad = - v_i \left( v_i - v_j - (z_{ij}S_{ij}^H + z_{ij}^H S_{ij}) + |z_{ij}|^2 \ell_{ij} \right),
\label{eq:2x2rank1}
\end{align}
where the last equality follows from \eqref{eq:Kirchhoff.2d}.   Substituting
\eqref{eq:Kirchhoff.2c} into \eqref{eq:2x2rank1} yeilds
$[W_G]_{ii} [W_G]_{jj} = \left| [W_G]_{ij} \right|^2$.
This together with $[W_G]_{ii} \geq 0$ (from \eqref{eq:ginv.1}) means 
$W_G(i,j) \succeq 0$ and $\rank {W_G}(i,j) = 1$.

\noindent
{\em Step 4: Proof that $g(W_G) = x$.}
Steps 1--3 show that $W_G := g^{-1}(x) \in \FF_2$ and hence $g$ has an inverse.  
We now prove this inverse is $g^{-1}$ defined by \eqref{eq:ginv.1}--\eqref{eq:ginv.2}. 
It is easy to see that \eqref{eq:g.1}--\eqref{eq:g.2} follow directly from \eqref{eq:ginv.1}--\eqref{eq:ginv.2}.  We hence are left to show that $W_G$
satisfies \eqref{eq:g.3}.   For each edge $i\rightarrow j\in \tilde E$ we have
from \eqref{eq:ginv.1}--\eqref{eq:ginv.2}
\begin{align*}
& |y_{ij}|^2 \left( [W_G]_{ii} + [W_G]_{jj} - [W_G]_{ij} - [W_G]_{ji} \right) \\
& \quad = |y_{ij}|^2 \left( v_i + v_j -  2\,  \Re(v_i - z_{ij}^H S_{ij}) \right) \\
& \quad = |y_{ij}|^2 \left( v_j -  v_i  + 2\,  \Re( z_{ij}^H S_{ij}) \right) \\ 
& \quad = \ell_{ij},
\end{align*}
where the last equality follows from \eqref{eq:Kirchhoff.2c}.
Hence $W_G$ satisfies \eqref{eq:g.3} and $g(W_G) = x$.
\end{IEEEproof}

We end this section with a visualization of Theorems \ref{thm:main}, \ref{thm:bfmMain} and \ref{thm:branch} in Figure \ref{fig:setRelations}. For any chordal extension $G_{ch}$ of graph $G$, the bus-injection model leads to three sets of problems $\PP_1, \PP_{ch}$, and $\PP_{2}$ and their corresponding relaxations $\RR_1, \RR_{ch}$ and $\RR_2$ respectively. 
The branch flow model leads to an equivalent OPF problem $\tilde \PP_2$, a nonconvex
relaxation $\tilde \RR_2^{nc}$ obtained by eliminating the voltage phase angles, and its
convex relaxation $\tilde \RR_2$.
The feasible sets of these problems, their relations, and the equivalence of the two
 models are explained in the caption of Figure \ref{fig:setRelations}.

%% file: numerics2.tex
\section{Numerics}
\label{sec:numerics}
We now illustrate the theory developed so far through simulations. 
First we visualize in Section \ref{sec:3bus} the feasible sets of OPF and their relaxations
for a simple 3-bus example from \cite{Lesieutre-2011-OPFSDP-Allerton}.
Next we report in Section \ref{sec:ieee} the running times and accuracies (in terms of exactness) 
of different relaxations on IEEE benchmark systems.
\begin{figure}[ht]
		\centering
    	    	\includegraphics[width=0.4\textwidth]{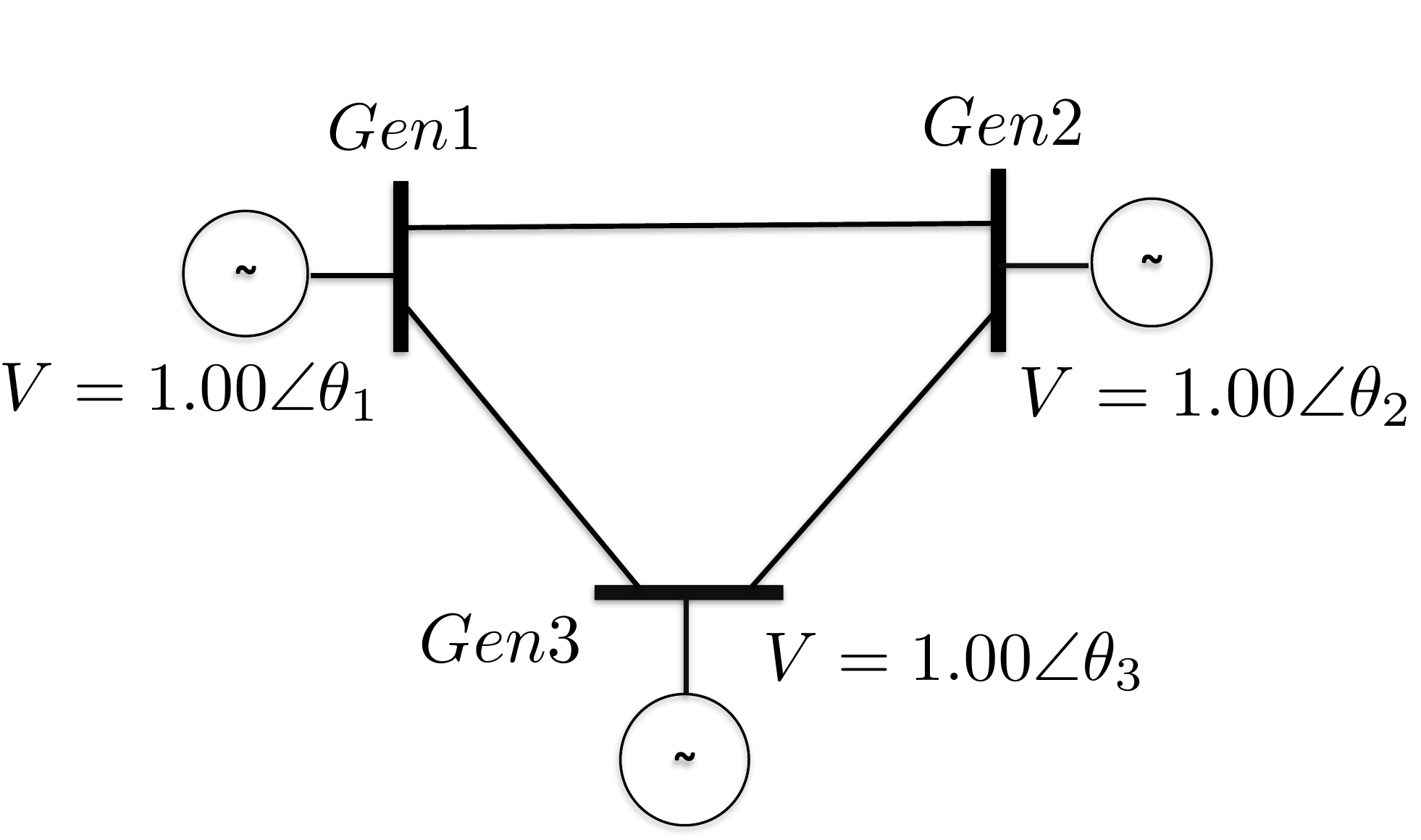}
    		\captionof{figure}{A 3-bus network from \cite{Lesieutre-2011-OPFSDP-Allerton}.}
    		\label{fig:3bus}
\end{figure}
\begin{table}[ht]
    \centering
    \begin{tabular}{| c | l | }\hline
      Parameter & Value \\ \hline
	$y_{11}$ & \ii 0.3750 \\
	$y_{22}$ & \ii 0.5 \\
	$y_{33}$ & \ii 0.5750 \ \\
	$y_{12}$ & 0.0517 - \ii 1.1087 \\
	$y_{13}$ & 0.1673 - \ii 1.5954 \\
	$y_{23}$ & 0.0444 - \ii 1.3319 \\ \hline
      \end{tabular}
      \captionof{table}{Admittances for the 3-bus network.}
          \label{table:case3}
\end{table}


\subsection{A 3-bus example}
\label{sec:3bus}


Consider the 3-bus example in Figure \ref{fig:3bus} taken from
\cite{Lesieutre-2011-OPFSDP-Allerton} (but we do not impose line limits)
with  line parameters in per units in Table \ref{table:case3}. 
Note that this network has shunt elements. 
For this example, $\PP_1$ is the same problem as $\PP_{ch}$  and 
$\RR_1$ is the same problem as $\RR_{ch}$.  Hence we will focus on the feasible sets
of $\PP_1$ (which is the same as that of $\PP_2$) and the feasible sets of $\RR_1$, $\RR_2$.
Each problem has a Hermitian $3 \times 3$ matrix $W$ as its variable. Recall that $s_j = p_j + \ii q_j$ is the complex power injection at node $j \in N$ and thus for each Hermitian matrix $W$, we have the following map:
$$ p_j(W) + \ii q_j(W) = W_{jj} \ y_{jj}  + \sum_{k: (j,k)\in E} \left( W_{jj} - W_{jk} \right) y_{jk}^H. $$

To visualize the various feasible sets, define the following set in 2 dimensions:
\begin{align}
\mathcal{A}_1 &:= \left\{ \left( p_1(W), p_2(W) \right) \ \vert \ W \in \FF_1, \right. \notag\\
& \qquad \quad \left. W_{11} = W_{22} = W_{33} = 1, p_3 (W) = -0.95 \right\}.
\label{eq:defA}
\end{align}

\begin{figure}[thb]
\centering
\subfigure[]{ { \scalebox{0.42}{\includegraphics*{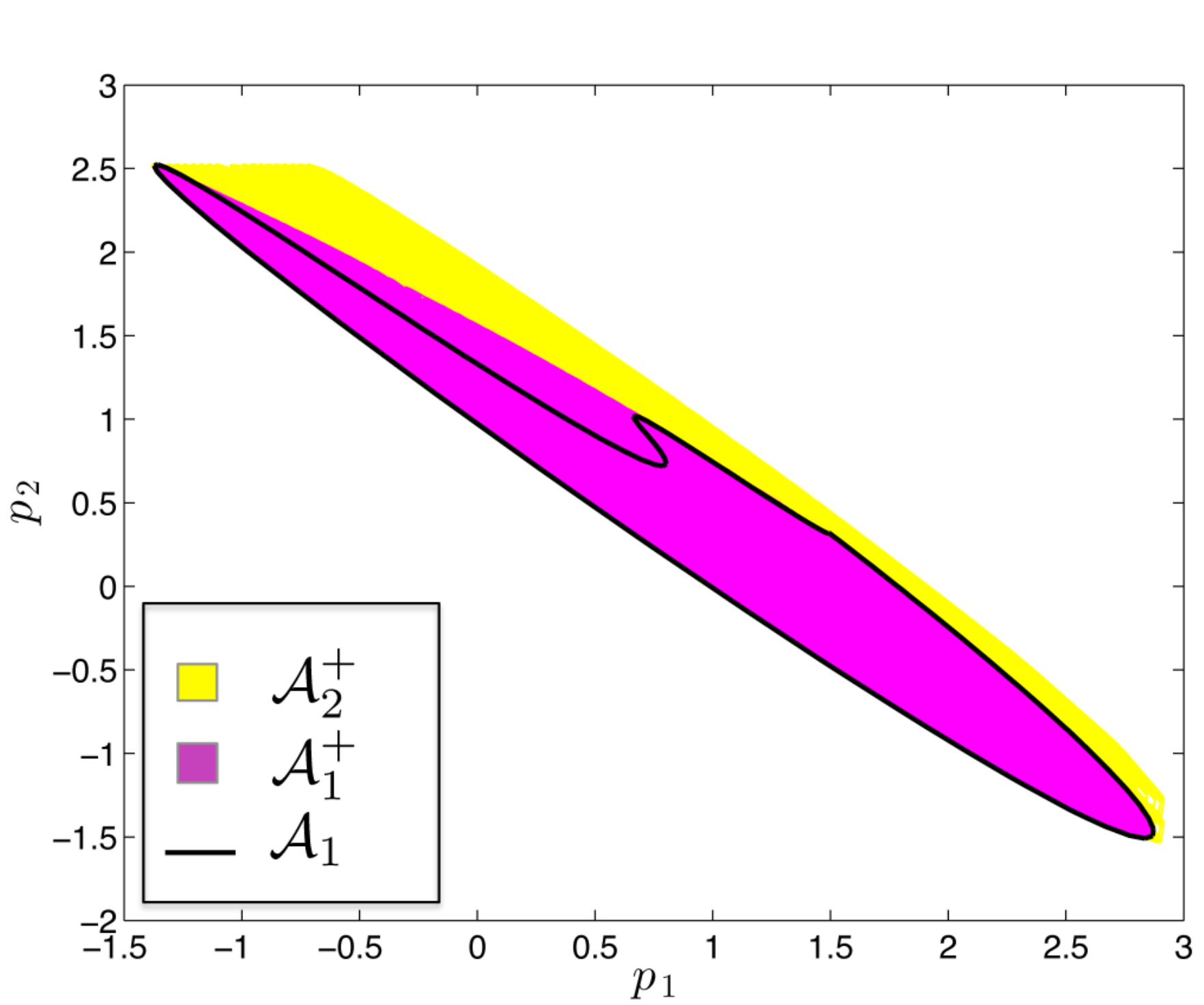}}} \label{fig:p1p2_socp_sdp} }
\subfigure[]{ {\scalebox{0.42}{\includegraphics*{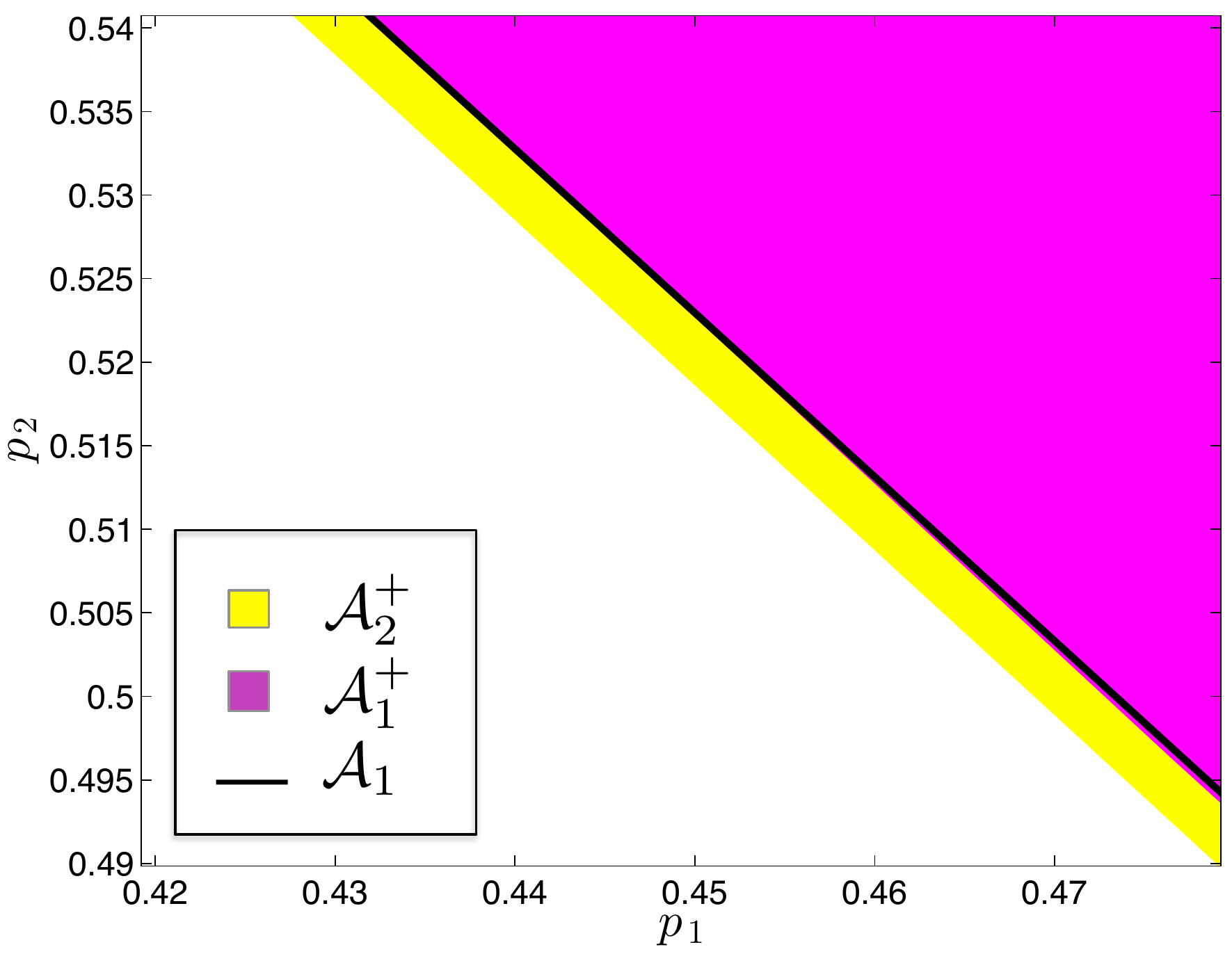}}} \label{fig:zoom}}
\caption{(a) Projections of feasible regions on $p_1 - p_2$ space for the 3-bus system in 
Figure \ref{fig:3bus}. \\(b) Zoomed-in Pareto fronts of these sets.}
\end{figure}

This is the projection of the feasible set of $\PP_1$ on the $p_1-p_2$ plane. Similarly, define the sets $\mathcal{A}_1^+$ and $\mathcal{A}_2^+$ where the Hermitian matrix $W$ is restricted to be in $\FF_1^+$ and $\FF_2^+$, respectively. We plot $\mathcal{A}_1$, $\mathcal{A}_1^+$ and $\mathcal{A}_2^+$ in Figure \ref{fig:p1p2_socp_sdp}. 
It illustrates the relationship among the sets in Figure \ref{fig:setRelations}, i.e., $\FF_1 \subseteq \FF_1^+ \subseteq \FF_2^+$. 
From Figure  \ref{fig:p1p2_socp_sdp},  $\mathcal{A}_1$ is non-convex 
while $\mathcal{A}_1^+$ and $\mathcal{A}_2^+$ are convex.  Since 
 $W \to (p_1(W), p_2(W))$ is a linear map, this confirms that 
  $\FF_1$ is non-convex while $\FF_1^+$ and $\FF_2^+$ are convex. 
  To investigate the exactness of relaxations, consider the Pareto fronts of the various sets (magnified in Figure \ref{fig:zoom}). 
The Pareto front of $\mathcal{A}_1^+$ coincides with that of $\mathcal{A}_1$ and thus relaxation $\RR_1$ is exact; relaxation $\RR_2$, however, is not.\footnote{SDP here are exact while some of the simulations in \cite{Lesieutre-2011-OPFSDP-Allerton} are not exact because we do not impose line limits here.}

\begin{figure}[h]
\centering
\includegraphics[width=0.42\textwidth]{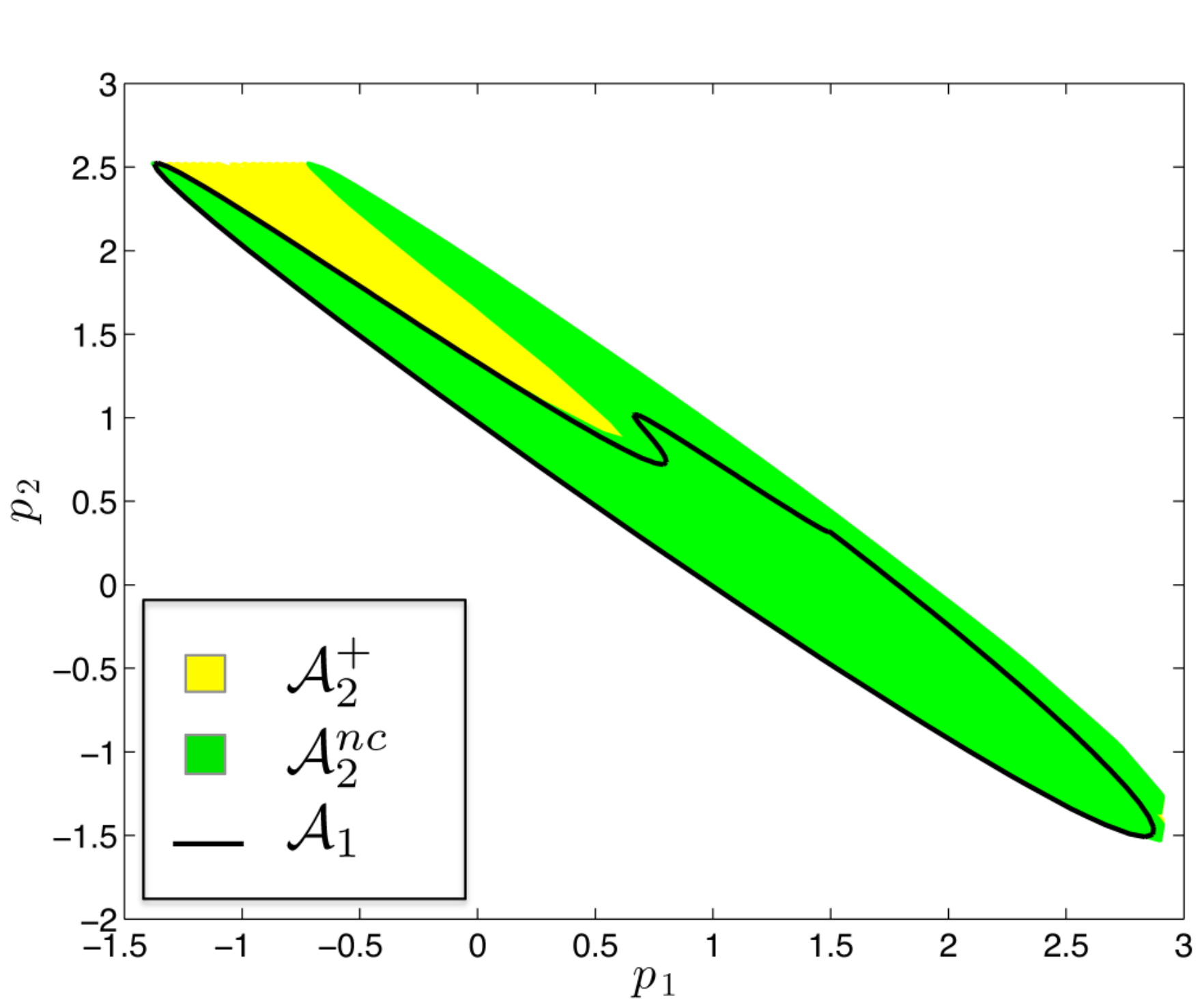}
\caption{Projections of feasible regions on $p_1 - p_2$ space for the 3-bus system in 
Figure \ref{fig:3bus}.}
\label{fig:p1p2_socp_exact}
\centering
\subfigure[]{ { \scalebox{0.42}{\includegraphics*{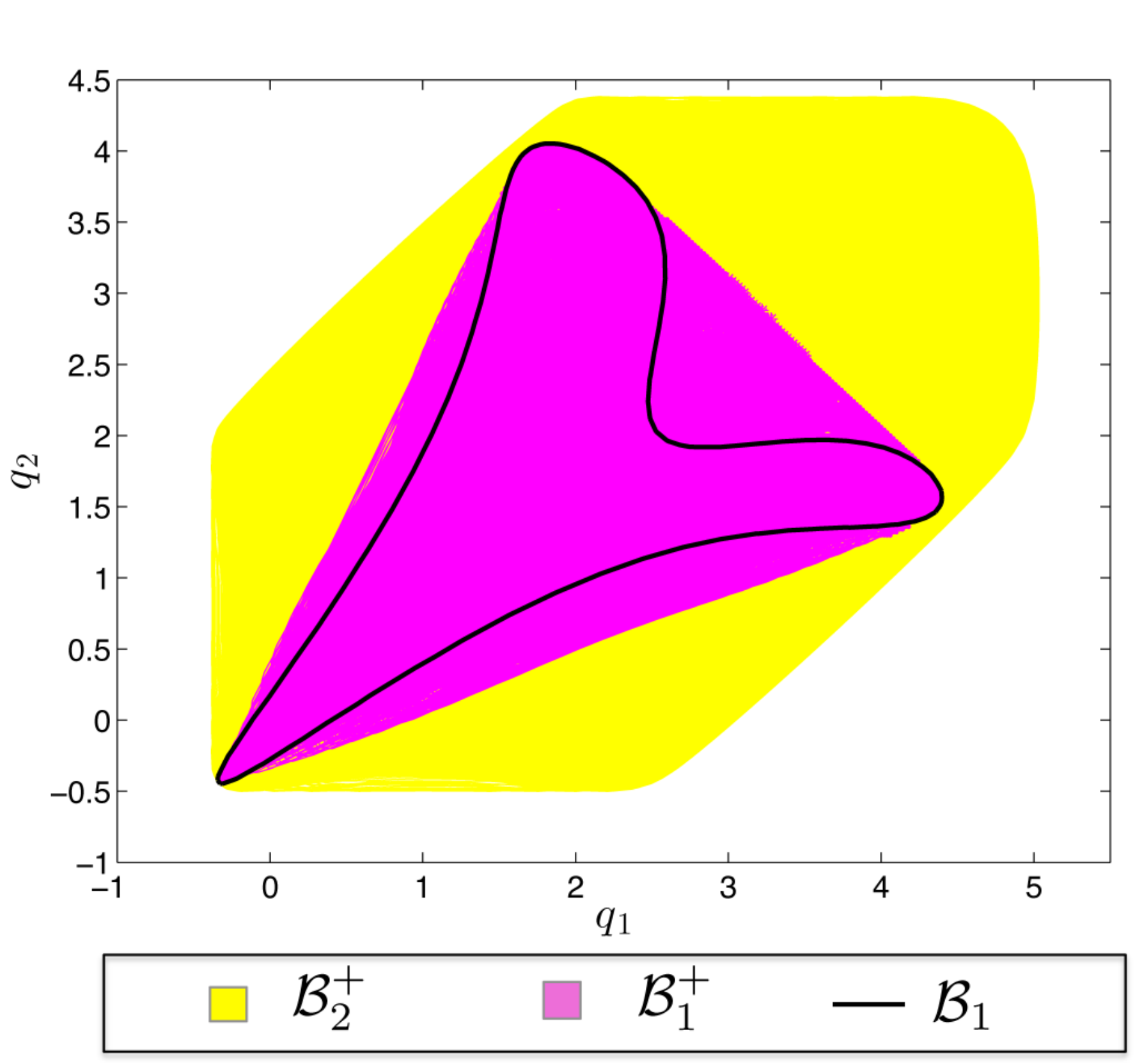}}} \label{fig:q1q2_socp_sdp} }
\subfigure[]{ {\scalebox{0.42}{\includegraphics*{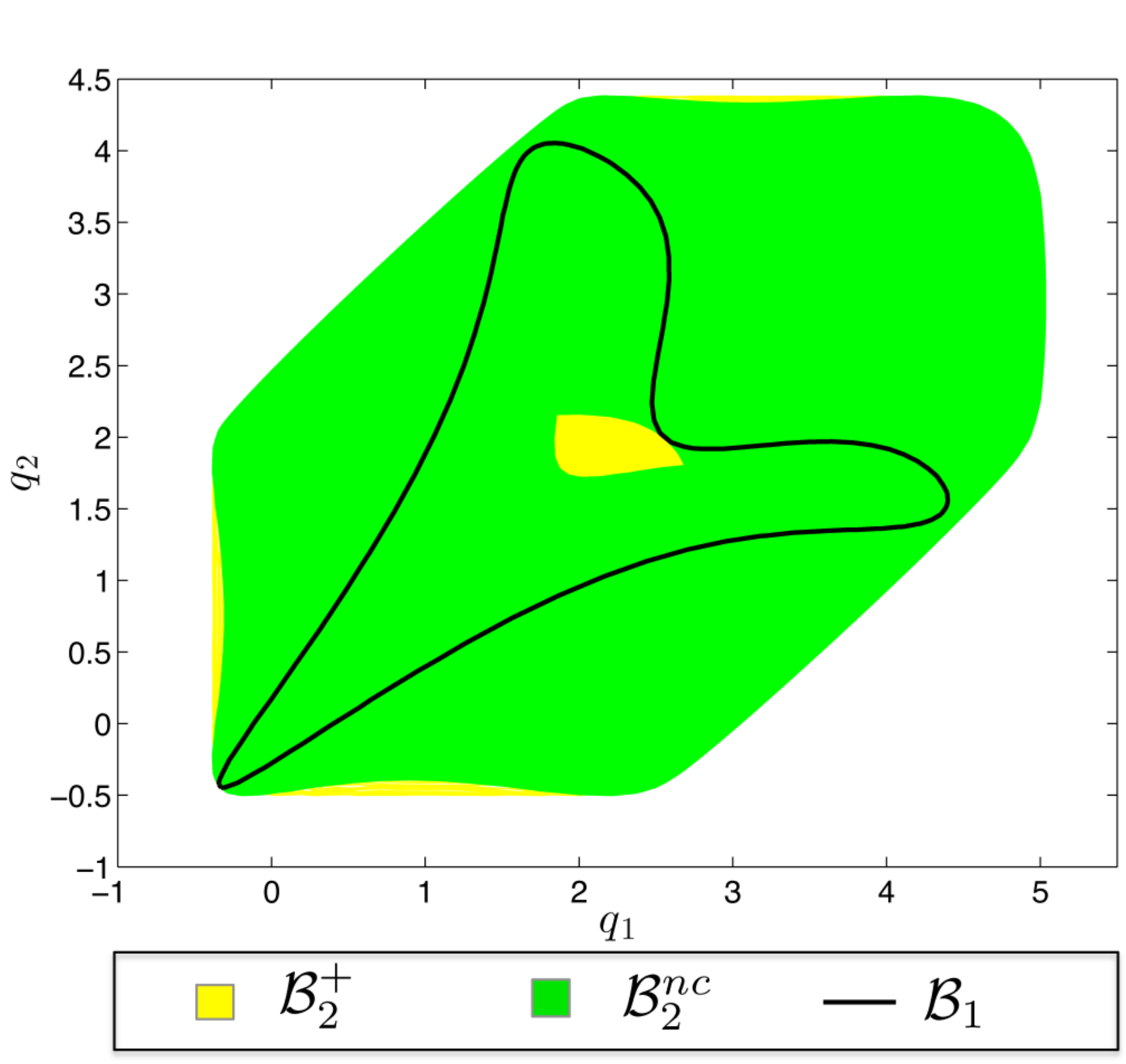}}} \label{fig:q1q2_socp_exact}}
\caption{Projections of feasible regions on $q_1 - q_2$ space for 3-bus system in Figure \ref{fig:3bus}.}
\end{figure}

Consider the set $\FF_{2}^{nc}$ defined in \eqref{eq:W2nc} that is equivalent to
$\mathbb X_2^{nc}$.   For this example, $\FF_{2}^{nc}$ is the set of $3 \times 3$ matrices $W$ that satisfy \eqref{eq:opfW.1}-\eqref{eq:opfW.2}
and the submatrices $W(1,2), W(2,3), W(1,3)$ are psd rank-1. The full matrix $W$, however, 
may not be psd or rank-1. Extend the definition of $\mathcal{A}_1$ in \eqref{eq:defA} to define the set $\mathcal{A}_2^{nc}$ where the matrix $W$ is restricted to be in $\FF_2^{nc}$. In Figure \ref{fig:p1p2_socp_exact}, we plot $\mathcal{A}_2^{nc}$ along with $\mathcal{A}_2^{+}$ and $\mathcal{A}$.
This equivalently illustrates the relation of the sets on the right in Figure \ref{fig:setRelations}.

For the projections on the $q_1-q_2$ plane define the set
\begin{align*}\mathcal{B}_1 & := \left\{ \left( q_1(W), q_2(W) \right) \ \vert \ W \in \FF_1, \right. \notag\\
& \qquad \quad \left. W_{11} = W_{22} = W_{33} = 1, \ p_3 (W) = -0.95 \right\}.
\end{align*}
As before, extend the definitions to $\mathcal{B}_1^+$, $\mathcal{B}_2^+$, and $\mathcal{B}_2^{nc}$. 
We plot $\mathcal{B}_1$, $\mathcal{B}_1^+$ and $\mathcal{B}_2^+$ in Figure \ref{fig:q1q2_socp_sdp} and 
$\mathcal{B}_1$, $\mathcal{B}_{2}^{nc}$ and $\mathcal{B}_2^+$ in Figure \ref{fig:q1q2_socp_exact}. 
This plot illustrates that the set $\FF_{2}^{nc}$ is not simply connected (a set is said to be simply connected if any 2 
paths from one point to another can be continuously transformed, staying within the set).
Note that neither of $\mathcal{B}_1^+$ and $\mathcal{B}_{2}^{nc}$ contains the other.



\begin{table*}[bt]
\centering
\begin{tabular} {|c|c|c|c|c|c|c|}
\hline
Test case&\multicolumn{2}{c|}{Objective value} & \multicolumn{3}{c|}{Running times} & Lambda ratio \\
\cline{2-6}
& $R_{1}, R_{ch}$ & $R_{2}$ & $R_{1}$ & $R_{ch}$ & $R_{2}$ & \\
\hline
9 bus & 5297.4 & 5297.4 & 0.2 & 0.2 & 0.2 &$1.15 \times 10^{-9}$
\\
14 bus& 8081.7 & 8075.3 &0.2 & 0.2 & 0.2 &$8.69 \times 10^{-9}$
\\
30 bus & 574.5 & 573.6 & 0.4&0.3 &0.3 &$1.67 \times 10^{-9}$
\\
39 bus & 41889.1 & 41881.5 & 0.7 &0.3 &0.3 &$1.02 \times 10^{-10}$
\\
57 bus & 41738.3 & 41712.0 &1.3 & 0.5&0.3 &$3.98 \times 10^{-9}$
\\
118 bus& 129668.6 & 129372.4 &6.9 &0.7 &0.6 &$2.16 \times 10^{-10}$
\\
300 bus & 720031.0 & 719006.5 &109.4 &2.9&1.8 & $1.26 \times 10^{-4}$ 
\\
2383wp bus & 1840270 & 1789500.0 & - & 1005.6 & 155.3 & median = $3.33 \times 10^{-5}$, max =$0.0034$. 
\\
\hline
\end{tabular}
\caption{Performance comparison of relaxation techniques for IEEE benchmark systems.}
\label{table:compare_ieee}
\end{table*}

\subsection{IEEE benchmark systems}
\label{sec:ieee}
For IEEE benchmark systems \cite{UW_data, Zimmerman09}, we solve $\RR_1$, $\RR_{2}$ and $\RR_{ch}$ in MATLAB using CVX \cite{cvx2012} with the solver SeDuMi \cite{sedumi} after some minor modifications to the resistances on some lines \cite{Lavaei2012}\footnote{A resistance of $10^{-5}$ p.u. is added to lines with zero resistance.}. The objective values and running times are presented in Table \ref{table:compare_ieee}. 
The problems $\RR_1$ and $\RR_{ch}$ have the same optimal objective  value, 
i.e., $r_1^* = r_{ch}^*$, as predicted by Theorem \ref{thm:main}. We also report the ratios of the first two eigenvalues of the optimal $W^*$ in $\RR_1$\footnote{For the 2383-bus system, we only run $\RR_{ch}$. For the optimal $G_{ch}$-partial matrix $W_{ch}^*$, we report the maximum and the median of the non-zero ratios of the first and second eigenvalues of $W_{ch}^*(C)$ over all cliques $C$ in $G_{ch}$.}; 
for most cases, it is small indicating that the relaxation is exact. 
The optimal objective value of $\RR_2$ is lower ($r_2^* < r_1^*$), indicating that
the optimum of the SOCP relaxation that is computed is not feasible for $\PP_1$.
As Table \ref{table:compare_ieee} shows, $\RR_{ch}$ is much faster than $\RR_1$ for large networks.
The chordal extensions of the graphs are computed \emph{a priori} for each case \cite{klerk2010}. 
$\RR_2$ is faster than both $\RR_1$ and $\RR_{ch}$, but yields an infeasible solution for 
most IEEE benchmark systems considered.